\documentclass[preprint]{aastex}
\begin{document}

\title{Techniques and Review of Absolute Flux Calibration from the
  Ultraviolet to the Mid-Infrared}

\author{Ralph~C.\ Bohlin, Karl~D.\ Gordon\altaffilmark{1}, and P.--E. Tremblay}
\affil{{Space Telescope Science Institute, 3700 San Martin Drive,
Baltimore,  MD 21218, USA}
\altaffiltext{1}{also Sterrenkundig Observatorium, Universiteit Gent,
              Gent, Belgium}}

\begin{abstract}  

The measurement of precise absolute fluxes for stellar sources has been pursued
with increased vigor since the discovery of the dark energy and the realization
that its detailed understanding requires accurate spectral energy distributions
(SEDs) of redshifted Ia supernovae in the rest frame. The flux distributions of
spectrophotometric standard stars were initially derived from the comparison of
stars to laboratory sources of known flux but are now mostly based on calculated
model atmospheres. For example, pure hydrogen white dwarf (WD) models provide
the basis for the HST CALSPEC archive of flux standards. The basic equations for
quantitative spectrophotometry and photometry are explained in detail. Several
historical lab based flux calibrations are reviewed; and the SEDs of stars in
the major on-line astronomical databases are compared to the CALSPEC reference
standard spectrophotometry. There is good evidence that relative fluxes from the
visible to the near-IR wavelength of $\sim$2.5~\micron\ are currently accurate
to 1\% for the primary reference standards; and new comparisons with lab flux
standards show promise for improving that precision.

\end{abstract}

\keywords{stars: atmospheres --- stars: fundamental parameters
--- techniques: spectroscopic}

\section{Introduction}

Flux values in physical units for astronomical objects are required to make
comparisons to physical models (e.g. Kent et al. 2009).  Such comparisons are
done regularly for observations at all astronomical distances; from Solar System
objects to individual stars to nebulae to entire galaxies. The need for a more
precise flux calibration has recently been highlighted by the fact that the
uncertainties in the flux distribution of stellar standards are the dominant
systematic error in measuring relative fluxes of redshifted supernovae Ia's and,
thus, in determining the nature of the dark energy that is driving the observed
accelerating cosmic expansion (Sullivan et al. 2011). A more detailed
understanding of dark energy requires a precise and accurate comparison between
the fluxes of distant and nearby supernovae in the rest frame. Quantitative
descriptions of dark energy in terms of the Einstein equations of general
relativity are significantly improved when the relative flux with wavelength is
known to an accuracy of 1\% or better (Aldering et al. 2004). Proof of this
primary justification for precise flux standards is the award of the 2011 Nobel
Prize in Physics for discovery of dark energy, which uses the SN Ia technique
for mapping the history of cosmic expansion. The study of circumstellar dust
rings in the mid-infrared (Su et al. 2006) provides a second example of the need
for a more precise absolute flux calibration. In this case, model atmosphere
calculations provide the basis for absolute mid-IR stellar fluxes; and more
precise lab based absolute flux measurements of stellar standards are required
to verify and improve the model atmospheres. The models are fit to the visible
and near-infrared stellar SEDs, where there is little emission from the dust;
and the dust signature is the difference between the measured mid-IR flux
and the fitted model. Thus, the precision of the model flux distribution from
the visible through the mid-IR is critical.

Our ability to measure stellar brightness has progressed from ancient times when
the eye was the only detector, and the precision was about one magnitude, i.e. a
factor of 2.512. Photographic film provided a somewhat more accurate brightness
measures that improved again with bolometers and photomultiplier tubes in the
1950s and 1960s. Today, the state-of-the-art of modern two-dimensional
electronic detectors makes possible a precision of order 1\% in the
determination of the  physical energy distributions of stars. The precision of
laboratory reference standards has also improved to better than 1\% in
absolute physical flux (Brown et al. 2006); but there are no recent results on
comparison of modern lab standards to stars. Currently, the best
stellar flux standards are limited by the precision of model atmosphere
calculations of pure hydrogen WD stars. (See section 3.2.3.)

In principle, one non-variable star with a known spectral energy distribution
(SED) is sufficient to establish flux distributions for all stars in the sky.
Ideally, a spectrophotometer located above any atmospheric absorption could
measure the  brightness of any star relative to the one standard candle, as long
as dynamic range, linearity of response, and out-of-band stray light were not
serious issues. Deustua et al. (2013) also review  absolute flux calibrations
but with a much broader brush that covers the entire electromagnetic spectrum
from gamma waves to radio wavelengths. This work goes into greater detail on the
restricted range of ultraviolet (UV) to mid-infrared (mid-IR). Section 2 reviews
the propagation of point source stellar flux standards to absolute surface
brightness of diffuse objects and details the mathematical basis of instrumental
flux calibration. Section 3 reviews attempts to establish the fluxes  of a few
stars relative to laboratory flux standards, while Section 4 presents the
methodology for using model atmosphere calculations for standard star SEDs. In
particular, Section 4 explains the use of pure hydrogen white dwarf (WD) models
and \emph{Hubble Space Telescope} (HST) spectrophotometry to establish a set of
UV/optical/near-IR flux standards, which are available in the
CALSPEC\footnote{http://www.stsci.edu/hst/observatory/crds/calspec.html}
archive. In section 5, several other archives of stellar flux standards are
reviewed; and sample SEDs are compared to CALSPEC standards.

The CALSPEC database is the repository for the SEDs resulting from the work
described in this review. Included in CALSPEC are the models for the three
primary standard WDs, which are NLTE model atmosphere calculations as normalized
to an absolute flux level defined by a reconciliation of physics-based visible
and IR absolute measures. While SEDs from other sources appear in the CALSPEC
database, the most precise and internally consistent set of fluxes are HST
spectrophotometry mainly from the STIS and secondarily from the NICMOS
instruments that are calibrated with equally weighted observations of the three
primary WDs. Estimates for IR fluxes from model atmosphere grids are often
included with the STIS and NICMOS SEDs for wavelengths longward of the limits of
those instruments.

\section{Calibration Basics}

\subsection{Concept}

In principle, the generation of a network of stellar flux standards is just a
simple matter of measuring the background-subtracted net signal $n$ in some
units like electrons per second from the program star and $N$ from a primary
standard of known flux F, where the same instrument is used for both stars and
where F is at the same spectral resolution as the measurements. The
flux $f$ of the program star is simply the ratio of the signals times the flux
of the primary standard \begin{equation}{f={F~ n\over N}}\end{equation} or
\begin{equation}f=S n~,\label{fsimp}\end{equation} where $S$ = $F$ / $N$ 
is the
instrumental sensitivity. This simple case is for a stable instrumental
configuration with a linear response. Stability means that repeated observations
produce the same response, while linearity implies that the count rate is
directly proportional to the physical flux $F$, i.e. the ratio of flux to count
rate will be the same ratio of $F/N$ over the dynamic range of the system. There
is no restriction on the entrance slit or extraction aperture as long as the
same choice is made for both stars and the extracted count rate is repeatable
for both stars. The measured count rate can be in a certain aperture radius for
point source photometry or of a fixed height on the detector of an imaging
spectrophotometer. The sensitivity $S$ is really more properly an inverse
sensitivity, because a more sensitive instrument will have a higher count rate
for a source of the same flux.

One complication arises when the spectral resolutions of the flux standard
differs from the unknown star. A common example is a standard star with a
tabulated medium resolution SED. In the case of a spectrometer with a resolution
that is lower than the tabulated resolution of the standard, the calibration $S$
as a function of wavelength is defined as the convolution of the known SED, $F$,
with the instrumental line-spread function, LSF, divided by the count rate
spectrum, $N$, of the standard convolved with the LSF of the standard star
spectrum, which brings the numerator and denominator spectra of $S$ to the same
resolution and enables a pixel-by-pixel division of $F$ by $N$ after resampling
to the same wavelength scale. This procedure may fail for the case where a low
resolution standard star SED must be bootstrapped to a calibration of a much
higher resolution spectrometer where the sensitivity of the high resolution data
changes significantly over the resolution element of the standard SED. For
example, a single echelle order may have a variation in sensitivity by a factor
of 10 or more over a wavelength range covered by only one or a few resolution
elements of the flux standard. The calibration of high resolution spectroscopy
is greatly simplified, whenever a high-fidelity,  high-resolution model
atmosphere calculation is available. For broadband photometry, the average flux
of the standard over the bandpass must be calculated as detailed below, which is
straightforward, if the spectral resolution is much better than the band width.

\begin{figure} 
\centering 
\includegraphics*[height=6in]{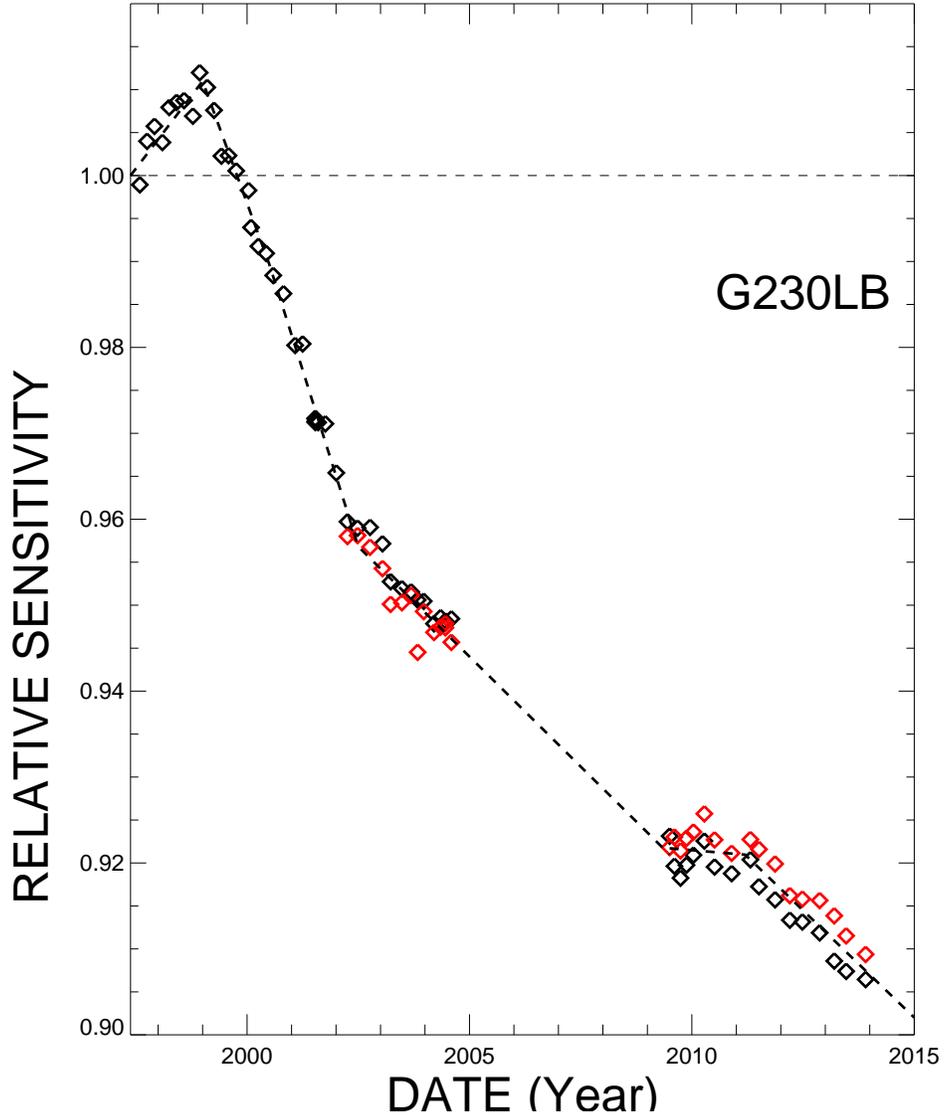}
\caption{\baselineskip=12pt
Changing sensitivity with time for the G230LB CCD mode of STIS on HST. Diamonds
are the average response to the monitoring standard AGK+81$^{\circ}$266 over the
2000--3000~\AA\ range. Black diamonds are for the center of the CCD detector,
while the red diamonds are for the E1 position near the readout amp. The dashed
lines through the data points are the piecewise linear fit to the changing
sensitivity, as normalized to unity at the beginning of the mission. All data
points are corrected for CTE losses. There is a data gap of almost five years
before STIS was repaired by a Servicing Mission in 2009. Presumably, the small 
sensitivity increase at the beginning of the mission is caused by evaporation of
contaminants from the optical surfaces.
\label{tchang}}\end{figure}

Other complications include non-linearity and changing instrumental response
with time or temperature. For CCD detectors, charge transfer efficiency (CTE)
during readout is not perfect and degrades with time, especially in the presence
of ionizing radiation which creates charge traps. A source near the readout
amplifier will be brighter than when the same star is placed farther away, which
creates a non-linearity that must be corrected in the flux calibration process.
A common non-linearity for pulse counting detectors is coincidence of pulses at
the higher count rates. Degradation of optical surfaces and detector quantum
efficiency over the long term must be tracked and accounted in the correction of
observed response to a reference epoch. For example, Figure~\ref{tchang} shows
the mean degradation for one UV mode of STIS since its installation in the HST
in 1997, while Anderson \& Bedin (2010) provide a good example of a CTE
correction. Because CTE and optical degradation both manifest as slow losses in
sensitivity, separate calibration programs are required to isolate and quantify
the two different effects.  

In the next section, $N_e$ in electrons s$^{-1}$ represents the instrumental
response corrected for all non-linearities, while corrections for temporal
instabilities can be applied at any stage to the sensitivity, the count rate, or
just the final fluxes.

\subsection{Definitions \& Equations for Spectrophotometry and Photometry}

\subsubsection{Photometry}

\subsubsubsection{Point Source}

The \emph{HST} method of flux calibration for filter photometry does not involve
color corrections or nominal wavelengths and is always defined in terms of the
photon weighted mean flux over the bandpass in wavelength units, where our flux
is in energy units (e.g. erg cm$^{-2}$ s$^{-1}$ \AA$^{-1}$)
\begin{equation}\langle F_{\lambda}\rangle={\int F_\lambda~\lambda~R~d\lambda
\over \int \lambda~R~d\lambda}=S_\lambda N_e\label{favl}\end{equation}  or in
frequency units (e.g. erg cm$^{-2}$ s$^{-1}$ Hz$^{-1}$) \begin{equation}\langle
F_{\nu}\rangle={\int F_\nu~\nu^{-1}~R~d\nu \over \int  \nu^{-1}~R~d\nu}
\label{favnu}=S_\nu N_e\end{equation} (Koornneef et~al. 1986, Rieke et al.
2008). $R$ is the system fractional throughput, i.e. the total system quantum
efficiency; $S_\lambda$ and $S_\nu$ are the instrumental sensitivities as a
function of wavelength and frequency, respectively; and the integrals are
computed over the full bandpass of the filter. The integration is done in photon
units ($F_{\lambda}\lambda$), because UV/optical/NIR detectors are generally
photon detection devices, rather than  total energy sensing bolometers, although
Bessell \& Murphy (2012) demonstrate that counting photons is equivalent to
integrating the energy. Some authors, (e.g. Cohen, et~al. 2003), define our
product $\lambda~R$ as their response function of the system. Also, see
Stritzinger et al. (2005). 

The instrumental count rate $N_e$ can be either measured in an infinite aperture
or calculated as \begin{equation}N_e=A\int{F_\nu\over{h\nu}}~R~d\nu={A\over hc}
\int{F_\lambda~\lambda~R~d\lambda}~, \label{ne}\end{equation} where $A$ is the
collecting area of the primary mirror, $h$ is Planck's constant, and $c$ is the
speed of light. The predicted throughput $R$ can be adjusted as needed to make
the predicted and measured count rates equal for observations of a stellar flux
standard. $N_e$ represents  the instrumental response after making any required
corrections for non-linearities and temporal changes. For crowded field
photometry, $N_e$ is often measured in small radius apertures and corrected for
the fractional enclosed energy. For example for the Advanced Camera for Surveys
(ACS) on HST, Bohlin (2012) tabulates the fractional enclosed energy for
isolated bright stellar images. 

Source independent instrumental sensitivities $S$ are defined by dividing the
mean flux by the detected electrons s$^{-1}$, $N_e$, in an infinite-radius
aperture. If $N_e$ from Equation (\ref{ne}) is substituted in Equations
(\ref{favl}-\ref{favnu}), \begin{equation}S_\lambda={hc \over
A\int{\lambda~R~d\lambda}} \label{pl}\end{equation} \begin{equation}S_\nu={h
\over A \int{\nu^{-1}~R~d\nu}}.  \label{pnu}\end{equation}   
For example, the
\emph{HST} standard flux units are normally per unit wavelength; and the constant
$S_\lambda$ appears in the headers of \emph{HST} photometric images with the
keyword name $photflam$. For NICMOS and WFC3, $S_\nu$ with the keyword name
$photfnu$ is also included in the headers. Other instrumental archives, e.g.
Spitzer, store calibrated images in units of surface brightness.

The \emph{HST} calibration constants are normally derived from the source
independent Equations (\ref{pl}--\ref{pnu}) after any required adjustments are
made to the $R$ estimated from the product of laboratory component QE
measurements. These adjustments are derived by making the measured $N_e(obs)$ in
an infinite aperture match the predicted $N_e(pred)$ calculated from Equation
(\ref{ne}). In practice, a radius of something like the 5.5\arcsec\ for ACS is
defined as "infinite" (Sirianni et~al. 2005, Bohlin 2012); and the primary pure
hydrogen WDs G191B2B, GD71, and GD153 are the preferred standards used for $F$
in Equation (\ref{ne}). In theory, models of the instrumental PSF could help
define the encircled energy for an infinite aperture. In the case of HST,
considerable effort has been expended on the Tiny Tim PSF modeling software.
However, the Tiny Tim user manual states: ``At short wavelengths, it may not be
possible to compute a PSF larger than 7\arcsec. Generally, the models are not
very good past a radius of $\sim$2\arcsec, due to the effects of scatter and
high-frequency aberrations'' (Krist and Hook 2004).

The reconciliation of laboratory component throughputs versus the truth of
standard stars is achieved by adjusting the normalization of the filter
throughput or even by changing the quantum efficiency, QE, as function of
wavelength for the detector or filter when sufficient information exists (e.g.
de Marchi, et~al. 2004, Bohlin 2012). Thus, information about individual
component throughputs, such as the telescope or detector QE, may be inferred
when reconciling sensitivities for several filters with different central
wavelengths.

To complement the above estimates of mean flux for stars imaged in a particular
filter, an associated wavelength is often useful. In addition to the nominal
wavelength $\lambda_o$ of Reach et al. (2005), other common definitions
are the {\em mean} and {\em effective} wavelengths.
\begin{equation}\lambda_\mathrm{mean}={\int{\lambda~R~d\lambda} \over \int{
~R~d\lambda}}\label{lmean}\end{equation}
\begin{equation}\lambda_\mathrm{eff}={\int{F_\lambda \lambda^2~R~d\lambda} \over
\int{F_\lambda \lambda ~R~d\lambda}} \label{leff}\end{equation} Perhaps, most
useful is the source independent \emph{pivot-wavelength} $\lambda_p$ and
associated \emph{pivot-frequency} $\nu_p$, where $\lambda_p \nu_p=c$ and 
$\langle F_{\lambda}\rangle~\lambda_p=\langle F_{\nu}\rangle~\nu_p$.
\begin{equation}\lambda_p=\sqrt{{c~\langle F_{\nu}\rangle}\over {\langle
F_\lambda\rangle}}= \sqrt{\int{\lambda~R~d\lambda}\over
\int{\lambda^{-1}~R~d\lambda}} \label{pivl}\end{equation}  These various
measures of the associated wavelengths for a filter are given by Koornneef et~al.
(1986).

Having calculated the source independent \emph{pivot-wavelength} $\lambda_p$,
Equation (\ref{pivl}) provides a convenient formula for calculating $S_\lambda$
from $S_\nu$ values.
\begin{equation}S_\lambda={c~S_\nu\over{\lambda_p}^2}\label{plam}\end{equation}

\subsubsubsection{Diffuse Source}

For the surface brightness of diffuse sources, i.e. the specific intensity or
radiance $I$, there are analogous equations for the instrumental calibrations,
where $\Omega_{pix}$ is the size of a pixel in steradians or arcsec$^2$. If
$N_I$ in electrons s$^{-1}$ represents the linearized instrumental response for
the count rate per pixel in a region of diffuse surface brightness, then
\begin{equation}\langle I_{\lambda}\rangle={\int I_\lambda~\lambda~R~d\lambda
\over \int \lambda~R~d\lambda}=C_\lambda~N_I\label{sb}\end{equation} or in
frequency units \begin{equation}\langle I_{\nu}\rangle={\int
I_\nu~\nu^{-1}~R~d\nu \over \int
\nu^{-1}~R~d\nu}=C_\nu~N_I~.\label{sbnu}\end{equation}  The calibration
constants for the specific intensity calibration are related to those for point
sources by \begin{equation}C={S \over \Omega_{pix}}~,\label{cdiff}\end{equation}
where $C$ and $S$ have either the $\lambda$ or the $\nu$ subscript. The reason
that $S$ for a point source and an infinite aperture is required can be
visualized by the following gedenken experiment. Consider a field of point
source stars with flux $F$ at the same spacing as the pixel grid of the
detector, which is observationally indistinguishable from a field of uniform
surface brightness. The total count rate N over all pixels for one isolated star
is $N=F/S$, while the very same count rate N, but in each pixel, is recorded for
the infinite field of point sources, because the contribution to the count rate
at any distance for the isolated star is contributed equally by the star located
at that same distance in the dense field. Thus, only a division by the solid
angle per pixel is required to convert the infinite aperture point source
calibration to a diffuse source calibration.

While diffuse source calibration is simple in principle, several practical
considerations limit the precision. Because the wings of the PSF may cover a
significant portion of the instrumental field-of-view, direct measurement of the
signal in an infinite aperture suffers from inaccuracies in the flat-field
correction and S/N issues in the far wings, even for a bright isolated star.
Alternatively, modeling of the PSF is also fraught with difficulties in
calculating the last few percent of the energy in the far wings, which arise
from diffraction, atmospheric blurring, ghosting, scattered light, etc. Sources
such as galaxies with a diameter comparable to the PSF size will require a
different calibration coefficient (Equation 14) than for an infinite diffuse
source such as the night sky or zodiacal background. Such small objects require
an effort to model the geometry of the source.

\subsubsection{Spectrophotometry}
\subsubsubsection{Point Source}

For the flux calibration of spectral data, the equations are analogous to those
for photometry but where the count rates, sensitivities, fluxes, and throughput
$R$ are
one-dimensional arrays ordered by increasing wavelength (or frequency). The
implicit assumption is that the instrumental sensitivity varies slowly enough
across any pixel so that the sensitivity of that pixel is well represented by
its average sensitivity. Thus, the integrals go away in
Equations~(\ref{pl})--(\ref{pnu}), i.e.
\begin{equation}F_{\lambda}=S_\lambda~N_e={{hc \over 
{A~\lambda~R~d_\lambda}}N_e}\label{fspl}\end{equation} and 
\begin{equation}F_{\nu}=S_\nu~N_e={{h~\nu \over 
{A~R~d_\nu}}N_e}~,\label{fspnu}\end{equation}  where $d_\lambda$ and $d_\nu$ are
the spectral dispersions in \AA\ per pixel or Hz per pixel, respectively.
The product $A~R$ is known as the effective area of the instrument. $S$
can be derived directly from an observation of a known flux standard $F$ divided
by the observed count rate spectrum $N_e$, where $F$ is binned to the 
resolution of the observed $N_e$, as discussed above in section 2.1. In
practice, $N_e$ for an infinitely high extraction width is $N(h)$ / $\epsilon(h)$,
where the correction $\epsilon(h)$ for $N(h)$ from a spectral extraction $h$
pixels high is derived from high signal spectral images.

\subsubsubsection{Diffuse Source}

A specific intensity calibration for spectra is analogous to
Equation~(\ref{cdiff}), except that the uniform source of surface brightness
must be limited by a slit of width W in arcsec in the  dispersion direction.
\begin{equation}C={S \over {m~W}}~,\label{cdiffsp}\end{equation} where m is the plate
scale (\arcsec/px) in the direction perpendicular to the dispersion. 








\section{Comparison of Stars to Laboratory Flux Standards}

Direct comparisons between stars and laboratory flux standards, such as standard
lamps, is complicated for ground-based telescopes by the need to place the lamp at sufficient distance from
the telescope to simulate a point source and by the differential atmospheric
absorption between the standard lamp and the star light. The use of standard,
calibrated detectors as the basis for absolute fluxes is hampered by the
faintness of the starlight and by the need to establish the throughput of a
telescope plus detector system for a collimated, uniform beam that simulates
stellar illumination. Making these sorts of comparisons in the Earth's
atmosphere is further complicated by the need to know the atmospheric
transmission as a function of wavelength and time. The remainder of this section
summarizes attempts to establish stellar fluxes above the atmosphere with
instrumentation that has been calibrated with respect to lab flux standards.
Whenever possible, these laboratory pedigreed results are compared with the
HST/STIS CALSPEC, WD based fluxes, which are described in Section 4.

\subsection{Ultraviolet below 3300~\AA}

At far-UV wavelengths, the atmospheric opacity is so large that any attempts to
measure stellar fluxes must be done from above the atmosphere. Even at balloon
altitudes the absorption by oxygen is overwhelming below 2000~\AA. Thus,
sounding rockets and satellites are the vehicles used in the far-UV. Although
observing times are only a few minutes, a sounding rocket has the advantage that
the pre-flight calibration can be confirmed post-flight, as long as the
contamination control for re-entry and landing is adequate. Early results
assumed that the visible fluorescence of fresh sodium salicylate is constant
with wavelength in the UV (e.g. Opal 1968, Stecher 1970). At Ly$\alpha$, the
absolute flux was referenced to sealed NO ionization cells. Later, more
sophisticated measures of absolute flux became available, as discussed below.

This section is mainly a historical account of early attempts to compare UV
starlight to physics based, laboratory flux standards. The ordering is roughly
by time, where Figures~\ref{figoao}--\ref{figsol}
demonstrate improving agreement of the measured UV fluxes with the CALSPEC
system from the $\sim$10\% level in the 1970's to $\sim$3\% for the SOLSTICE
fluxes of Snow et al. (2012). Both the available lab standards and the
instrumental techniques have improved over the last five decades, which suggests
that a precision of 1\% is currently possible.

\subsubsection{Rockets}

\begin{figure}
\centering
\includegraphics*[height=5.6in,trim=25 0 0 0]{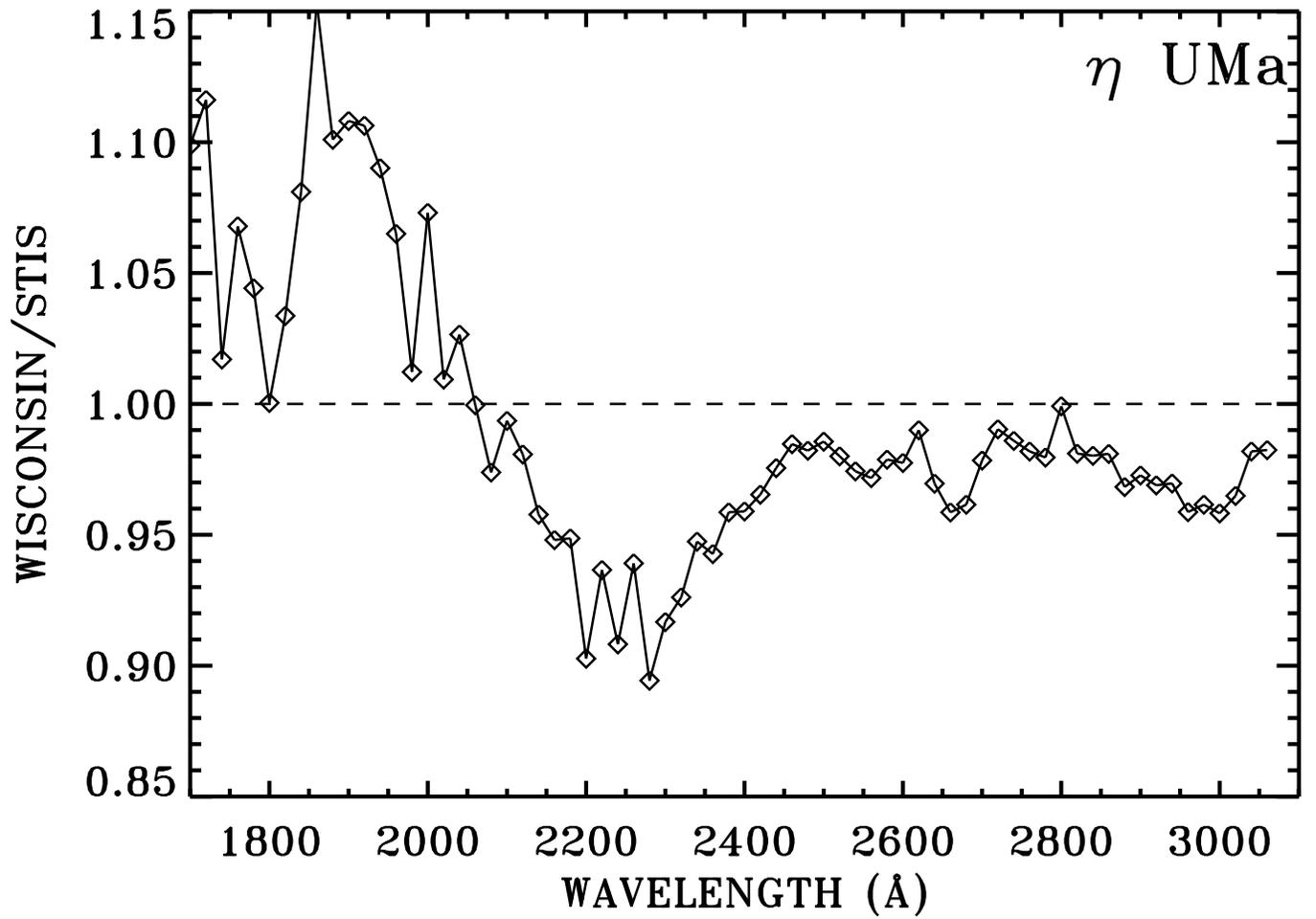}
\caption{\baselineskip=12pt
Ratio of OAO-2 fluxes based on a synchrotron calibration to the
baseline STIS fluxes for $\eta$ UMa.
\label{figoao}} \end{figure}

At the University of Wisconsin, an early sounding rocket program (Bless et al.
1976) provided stellar standards for the flux calibration of the Orbiting
Astronomical Observatory (OAO-2). In the 1370--2920~\AA\ wavelength range, the
known spectral energy distribution of 240~MeV electrons in the Wisconsin
Synchrotron storage ring established the sensitivity of the seven photometers
that comprised the rocket payload. The absolute flux of the synchrotron beam was
determined by basic physics and the number of circulating electrons. As the beam
degraded from an initial strength of about 50 electrons, the electron count was
determined by the incremental losses in signal strength.
Other factors entering the calculation were the transmission of the MgF$_2$
window on the storage ring port and the geometry of the illumination. A complete
recalibration was done after the flight and recovery of the payload. Three
stars, $\alpha$ Vir, $\eta$ UMa, and $\alpha$ Leo, were observed; and their
measured fluxes were used to update the calibration of  the spectrometer on
OAO-2. These absolute physical fluxes that are based on the physics of
synchrotron light emission have estimated uncertainties of 10\% shortward of
2000~\AA\ and 5\% longward of 2000~\AA. Over the wavelength range in common,
Figure~\ref{figoao} compares the resulting OAO-2 flux distribution for
$\eta$~UMa  with the modern HST/STIS baseline WD system of absolute fluxes from
the CALSPEC database.  

At the University of Colorado, another rocket program based its measured fluxes
on standards available from the National Institute of Standards and Technology
(NIST, but was NBS at the time). Bohlin et al. (1974) used a NBS pedigreed
standard photodiode detector for the 1164--2385~\AA\ range and a tungsten lamp
for 2250--3400~\AA\ to calibrate their spectrometer payload with its
$\sim$20~\AA\ spectral resolution. Because of the expected faint stellar signals
and the need to calibrate the spectrometer at these low levels, the NBS diode
calibration was first transferred to a photomultiplier tube, which was used to
scan the collimated, monochromatic input beam in the vacuum laboratory
calibration chamber. Every transfer or correction of a NBS pedigree is
associated with some added uncertainty, which degrades the precision of the
measured stellar fluxes.

The Colorado program used a standard tungsten ribbon light source at the longer
wavelengths; but such a hot lamp drawing 38.722 amps must be operated in air to
maintain proper convective cooling. A precision aperture in front of the
tungsten ribbon defined an effective point source when the lamp was placed at a
distance of 73~m from the flight spectrophotometer. However, this laboratory
arrangement required a correction for the attenuation by the atmosphere
(Strongylis \& Bohlin 1976), which caused the uncertainty to increase toward
shorter wavelengths. The lamp calibration was used for the final calibration of
the flight spectrometer down to 2400~\AA, where the uncertainty was estimated to
be +18/-7\%. A check of the relative sensitivity vs. wavelength was provided by
the molecular branching-ratio technique from observations of CO, NO, and $N_{2}$
spectra in the laboratory vacuum chamber. 

\begin{figure}
\centering
\includegraphics*{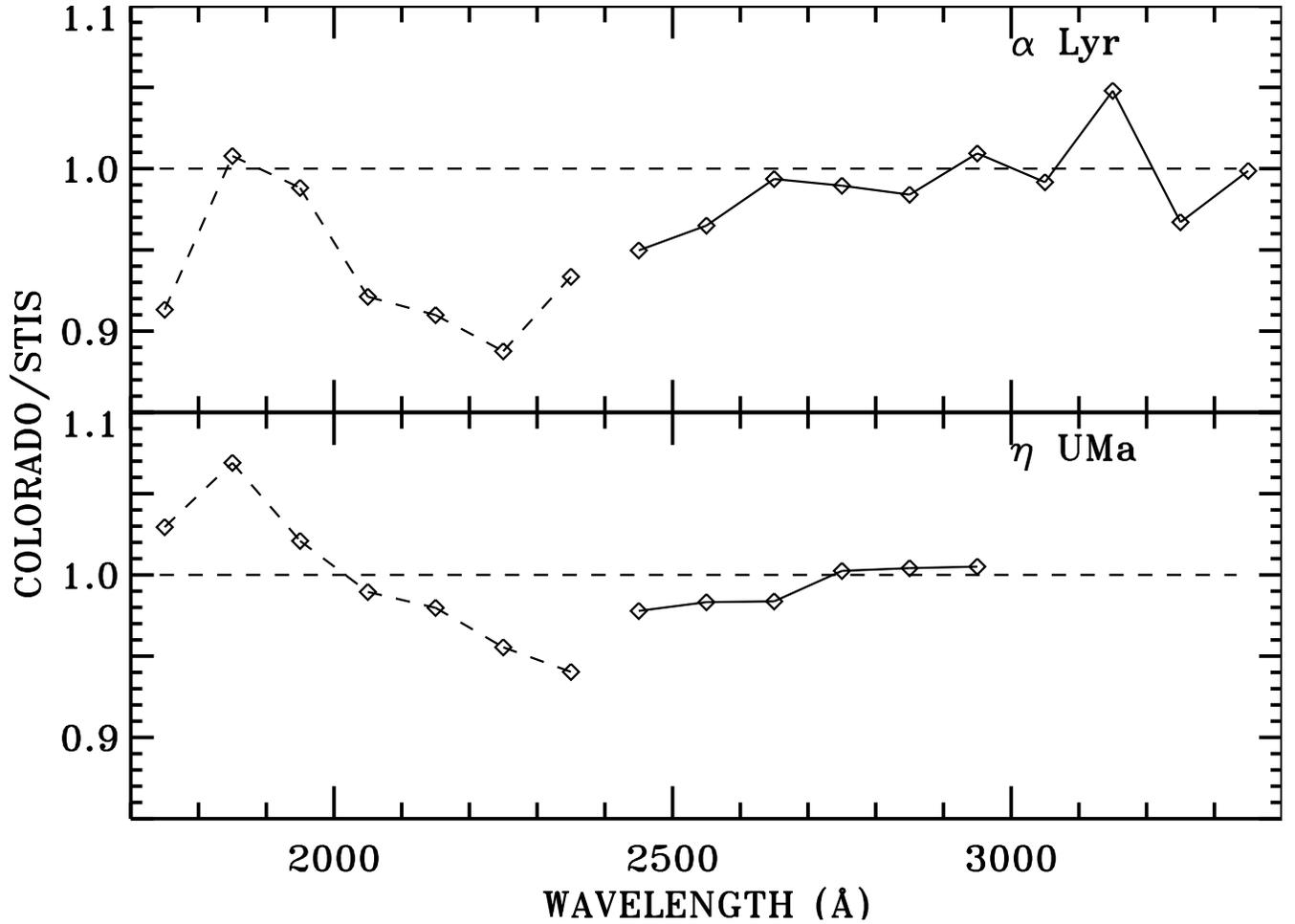}
\caption{\baselineskip=12pt
Ratio in 100~\AA\ bins of Strongylis \& Bohlin fluxes to the baseline STIS
fluxes. For the rocket fluxes below 2400~\AA\ (dashed line), the reference
standard is an NBS standard photodiode detector, while the solid line connects
fluxes that are referenced to an NBS standard light source. STIS fluxes for 
$\eta$ UMa are not available longward of 3000~\AA.
\label{figboh}} \end{figure}

Because of a failure in the short wavelength channel on the Colorado rocket,
results for only the 1700--3400~\AA\ range were obtained for the target stars
$\alpha$ Lyr, $\eta$ UMa, and $\zeta$ Oph (Strongylis \& Bohlin 1976). For two
stars, Figure~\ref{figboh} compares those measured fluxes to the baseline CALSPEC
database over their available common wavelength ranges in 100~\AA\ bands. 

\subsubsection{Manned Space Flights}

Manned space flights offer opportunities for longer integration times but have
the complications of preventing contamination from human effluents and
longer times from delivery to return of the flight instrument to the
laboratory calibration chamber. 

\subsubsubsection{Apollo 17}

Henry et al. (1975) flew an UV spectrometer on
the Apollo 17 mission to the Moon; and the flux calibration was referenced to an
NBS photodiode with an uncertainty estimate of 10\% over its 1180--1680~\AA\
wavelength range. Six stars were observed, including $\eta$ UMa, which lies
below the OAO-2 fluxes by as much as 28\% (Strongylis \& Bohlin 1976).

\subsubsubsection{Hopkins Ultraviolet Telescope}

\begin{figure}
\centering
\includegraphics*{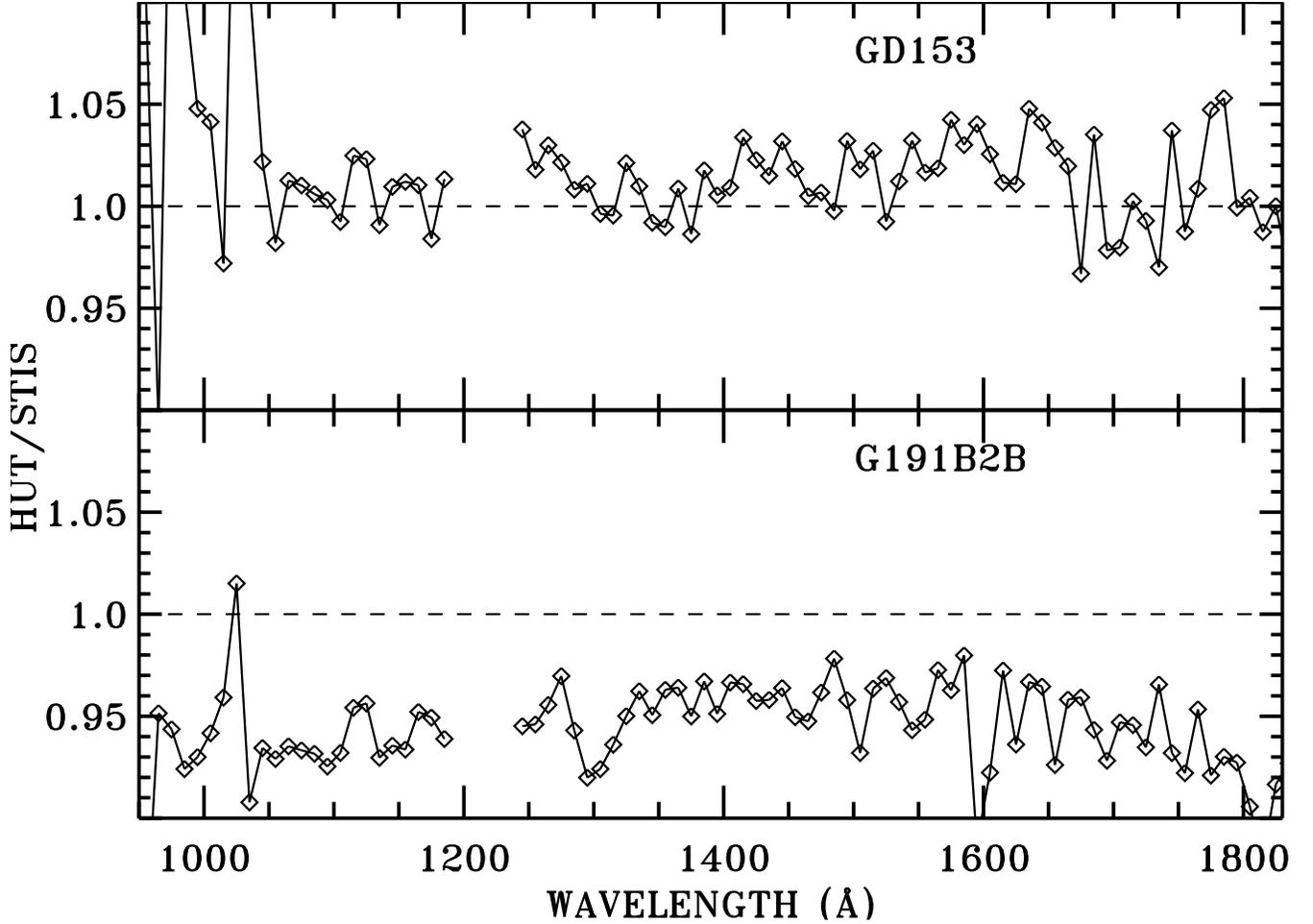}
\caption{\baselineskip=12pt
Ratio with a bin size of $\sim$10~\AA\ of HUT fluxes to the pure hydrogen WD
models for two of the HST primary flux standards. The gap is where the HUT
spectra are contaminated by geocoronal Ly$\alpha$, while the 1302~\AA\ OI aiglow
has been clipped from the observations. The large scatter below 1050~\AA\ is
caused by a slight mismatch of the strong Lyman absorption lines between the
observations and the models. \label{fighut}} \end{figure}

The Hopkins Ultraviolet Telescope, HUT, (Kruk et al. 1997) detected light in the
830--1840~\AA\ range. While the final calibration and archival fluxes were based
on the modeled fluxes for the WD G191B2B, an independent lab calibration was
also done with reference to NIST calibrated photodiodes and to the NIST
synchrotron facility. Longward of 912~\AA, the WD and NIST lab calibrations
agree to 3\% rms with a worst-case difference of 7\% near 1350~\AA. On average
over the 912--1840~\AA\ range, the lab synchroton calibration differs from the
adopted WD calibration by less than 0.5\%. At 830~\AA, there is a 20\%
discrepancy between the two methods. Figure~\ref{fighut} illustrates ratios of
the archived WD fluxes to the current model spectra for two of these primary HST
standards.

\subsubsection{Satellites}

Spectrophotometry from space based satellites provides the best intercomparisons
among a diversity of objects. Most space observatories relied on in-flight
observations of standard stars and {\em not} on direct comparison to lab flux
standards for their recommended flux calibration. This group includes the
{\it International Ultraviolet Explorer} (IUE), the {\it Far Ultraviolet Spectroscopic
Explorer} (FUSE), and the {\it Hubble Space Telescope} (HST). One observatory,
{\it Copernicus} (OAO-3), obtained high resolution spectra from stars focussed on a
narrow entrance slit. This {\it Copernicus} data set is not photometric, i.e.
observations were not repeatable, because the part of the stellar
point spread function (PSF) falling
in the slit varied from one acquisition to the next.

In general, laboratory calibrations of space observatories are not done
"end-to-end", because the whole instrument package is too big for most vacuum
facilities. Furthermore, the time between any laboratory flux calibration and
flight operations is long, allowing many opportunities for contamination of the
optical surfaces by thin polymer films which can absorp significant fractions of
the UV light beam. Often, estimates of total throughput are computed from the
throughput, i.e. quantum efficiency (QE), of the individual components from
primary mirror to detector with a corresponding accumulation of uncertainties.
UV satellite observatories with catalogs of absolute fluxes that are referenced
to lab flux standards are TD1, ANS, and SOLSTICE.

\subsubsubsection{Belgian/UK Ultraviolet Sky Survey Telescope}

The {\it Belgian/UK Ultraviolet Sky Survey Telescope} (S~2/68) in the ESRO TD1
astronomical satellite does have a flux calibration traceable to laboratory
standards. Launched in 1972, the S~2/68 UV spectrometer package was
independently flux calibrated about three months before launch by groups in the
United Kingdom (UK) and Belgium against an absolute radiometric detector and a
blackbody source, respectively (Humphries et al. 1976). The flight package was
relatively small with only a  27.5~cm diameter primary mirror; and the coverage
is 1350--2550~\AA\ with a resolution of 35~\AA. There was also a photometer with
a bandpass centered at 2740~\AA. For the UK absolute calibration, the
photomulipier tube used to measure the irradiance (flux) input to the TD1 flight
instrument was calibrated against a thermopile with a pedigree traceable to the
National Physics Laboratory in London. A thermopile with gold-black coatings to
absorb the light can measure the total energy in a monochromatic beam but is far
less sensitive than a photomultiplier detector tube. Thus, the crux of the
absolute flux calibration consisted of a scheme to reduce the intensity of the
illuminating beam by a known geometric factor of $4 \times 10^{5}$ between the
illumination of the thermopile and the detector.

\begin{figure}
\centering
\includegraphics*{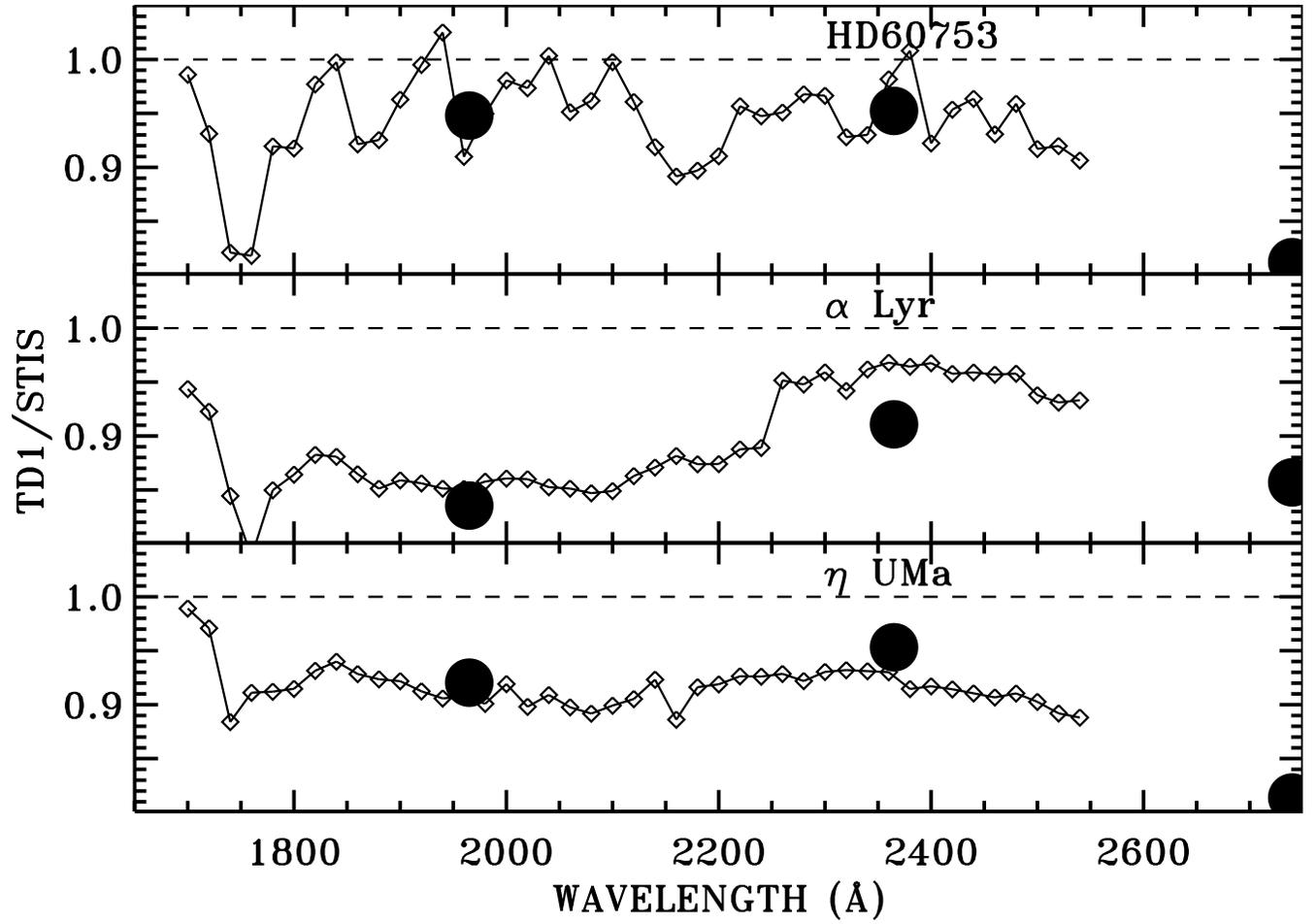}
\caption{\baselineskip=12pt
Ratio of TD1 fluxes to the baseline STIS fluxes with a bin size at the TD1
resolution of $\sim$35~\AA. The connected small diamonds are the early results
of  Jamar et al. (1976), while the large filled circles are from the on-line
catalog of Thompson et al. (1978). \label{figtd1}} \end{figure}

For the independent calibration in Belgium, the photomultiplier used to scan
the input beam to the flight instrument was referenced to a blackbody light
source where the radiant flux could be calculated from the Stefan-Boltzmann
relation for the operating temperature of $\sim$500~K. The blackbody radiation
at 6~$\micron$ illuminated an intermediate thermopile, which could also
measure the energy in a monochromatic UV beam. Attenuation at the
photomultipier was provided by a gold coated MgF$_2$ filter with a measured
transmission as a function of wavelength. This technique relies on the
absorption coeficients of the gold-black coated thermopile, which are 0.981
in the UV and 0.966 in the IR. 

The adopted final calibration is the average of the independent UK and Belguim
results, while the two separate calibrations differ from the mean by as much as
19\%. A whole-sky
catalog\footnote{http://webviz.u-strasbg.fr/viz-bin/VizieR?-source=II/59B} of
results is available on-line for 31215 stars (Thompson et al. 1978) down to
about V=10 for unreddened B stars in four passbands at 1565, 1965, 2365, and
2740~\AA. For a subset of the brighter stars, Jamar et al. (1976) present the
full spectra with a finer sampling interval. For three stars,
Figure~\ref{figtd1} compares these TD1 fluxes to the baseline STIS dataset over
their common wavelength range. Because these stars are too bright for the far-UV
STIS MAMA detectors, STIS fluxes have short wavelength limits of 1680~\AA.

\subsubsubsection{Astronomical Netherlands Satellite}

The {\it Astronomical Netherlands Satellite} (ANS) collected UV
photometry in five bands from 1550--3300~\AA; but the catalog (Wesselius et al.
1982)\footnote{http://cdsarc.u-strasbg.fr/viz-bin/Cat?cat=ans\&find=+} contains
only 3573 stars, and the documentation of the laboratory flux calibration
(Aalders et al. 1975) is not readily available. 

\subsubsubsection{Other UV Space Missions}

\begin{figure}
\centering
\includegraphics*[width=1.1\textwidth]{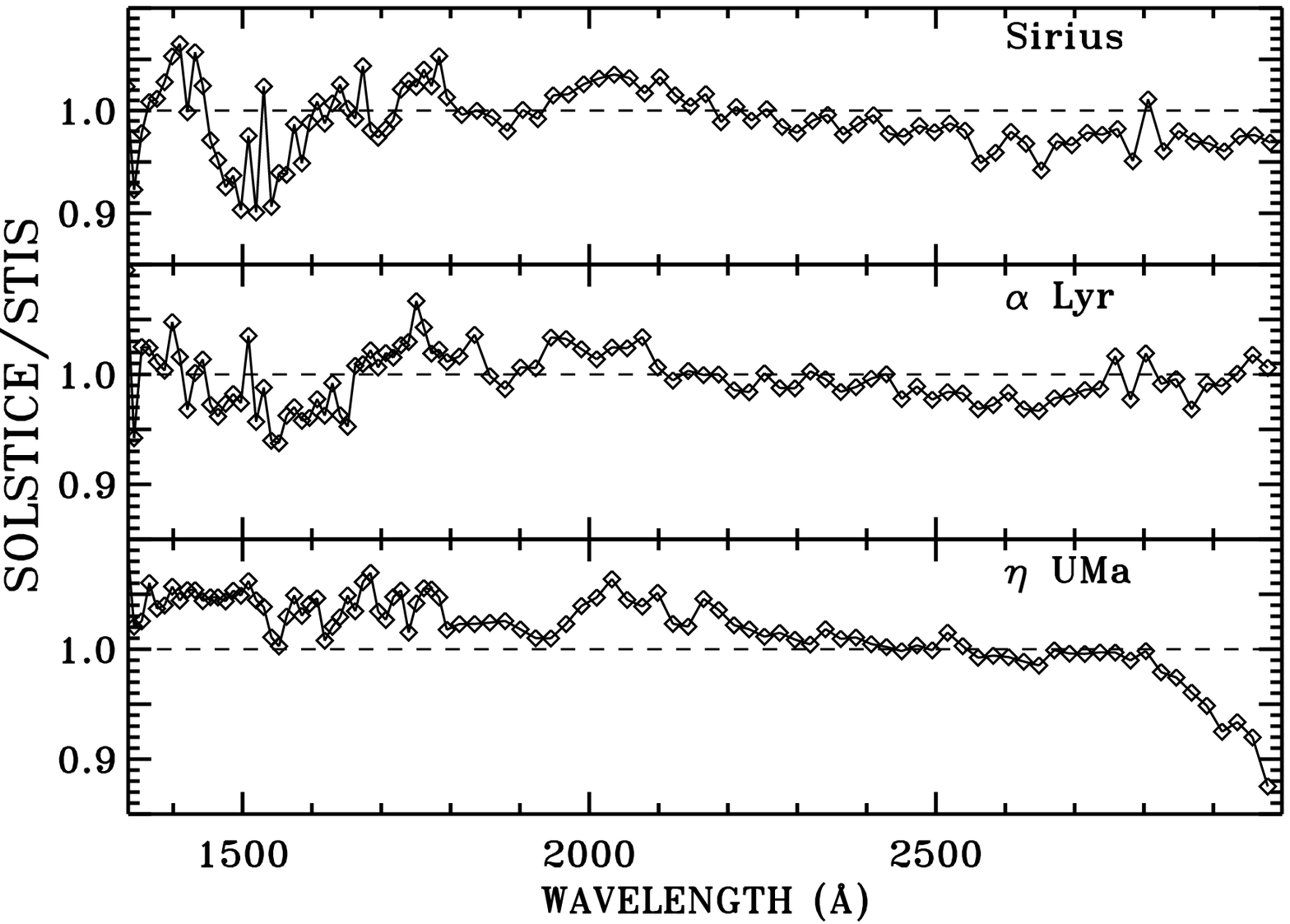}
\caption{\baselineskip=12pt
Ratio of SOLSTICE fluxes to the baseline STIS fluxes above 1700~\AA\ and IUE at
shorter wavelengths with a bin size at the SOLSTICE resolution. SOLSTICE fluxes
are referenced to a synchrotron flux standard, while the STIS+IUE fluxes are on
the WD HST system.\label{figsol}} \end{figure}

More recently, Snow et al. (2012) have published UV flux distributions for 18
bright A and B stars obtained by the SOLar-STellar Irradiance Comparison
Experiment (SOLSTICE) spectrometer on the SORCE spacecraft, which also measured
the solar flux. The absolute fluxes are tied to a synchrotron source, NIST SURF
III (Arp et al. 2000, McClintock et al. 2005) with coverage from 1150--3000~\AA\
at 11--22~\AA\ resolution. The estimated absolute accuracy is 3\%. 
Figure~\ref{figsol} compares the SOLSTICE fluxes for the three stars in common
with the STIS CALSPEC archive. However below 1700~\AA, the STIS has been
supplemented by IUE spectra matched in the overlap region, because the three
stars are too bright for the STIS FUV MAMA detector. Differences between the
synchrotron based SOLSTICE and the WD based fluxes rarely exceed  2$\sigma$,
i.e. 6\%.

\subsection{Visible 3300--10000~\AA}

Compare the discussion in this section to the review by Deustua et al. (2013).

\subsubsection{Historic}
 
Early work by Oke \& Schild (1970) and Hayes \& Latham (1975) on measuring the
absolute flux of the primary ground-based standard Vega has survived the test of
time. From 3300 to 10800~\AA, Oke \& Schild used a NBS pedigreed tungsten ribbon
lamp and two blackbody cavities operated at the melting point of copper to
measure the flux, F(5556), of $\alpha$ Lyr with a value of $3.36\times10^{-9}$~erg
cm$^{-2}$ s$^{-1}$~\AA$^{-1}$ $\pm$2\% at 5556~\AA. Hayes \& Latham improved the
atmospheric extinction corrections and utilized the earlier relative
measurements of Hayes (1970) and absolute fluxes in the 6800--10800~\AA\ range
from Hayes et al. (1975) to derive F(5556)=$3.45\times10^{-9}$~erg
cm$^{-2}$s$^{-1}$~\AA$^{-1}$ $\pm$2\%. The Hayes fluxes are based
entirely on copper melting-point blackbodies. Averaging with the  corrected
results of Oke \& Schild produced a best estimate for F(5556) of
$3.39\times10^{-9}$~erg cm$^{-2}$ s$^{-1}$~\AA$^{-1}$ $\pm$2\%. While Hayes \&
Latham present fluxes for the full 3300--10800~\AA\ range, the value at 5556~\AA\
is especially important for the WD flux system discussed in section 4.

\begin{figure}
\centering
\includegraphics*{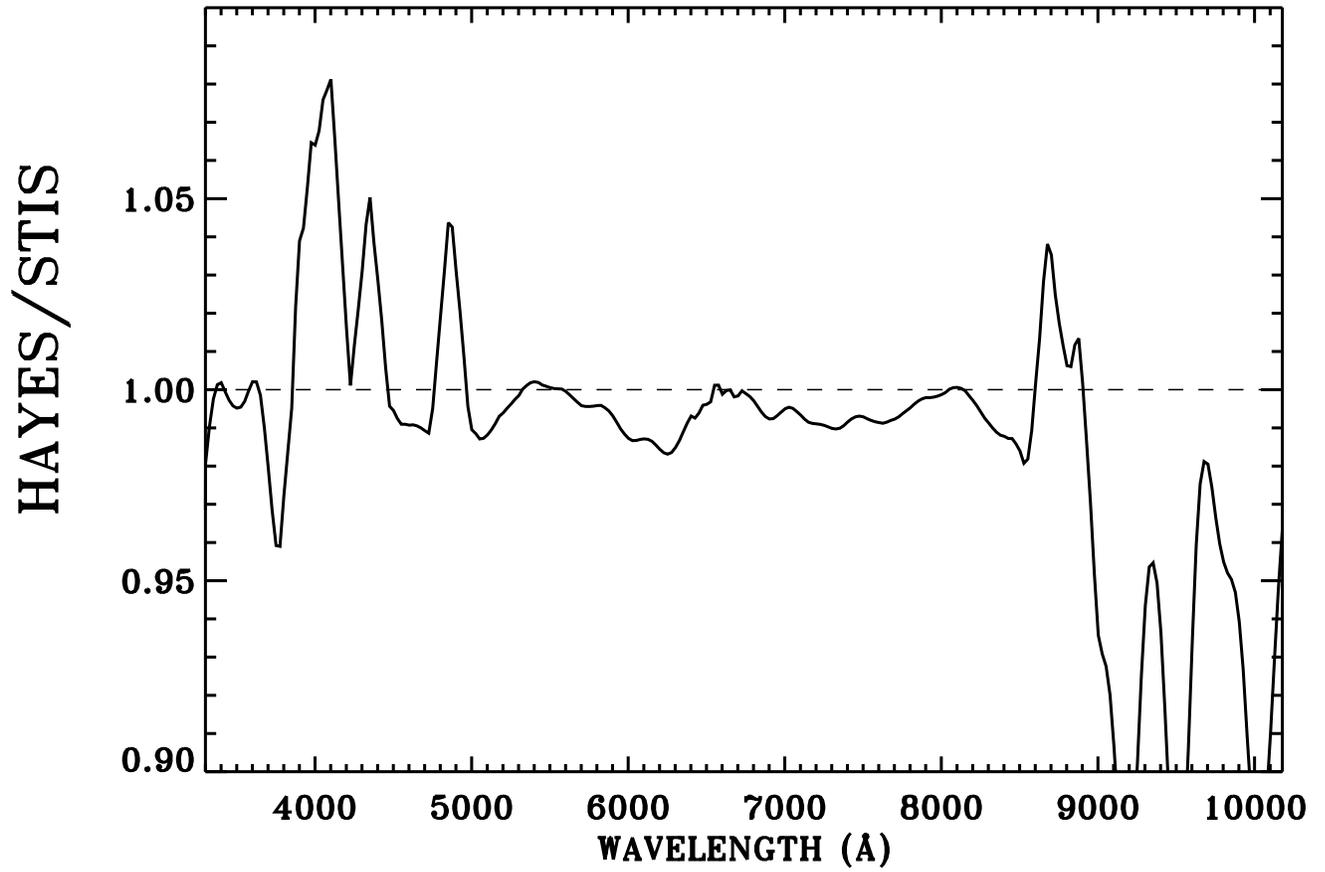}
\caption{\baselineskip=12pt
Smoothed ratio for Vega of the H85 to the baseline CALSPEC fluxes at the 25~\AA\
bin spacing of H85. \label{hayes}} \end{figure}

Hayes (1985, hereafter H85) reviewed the available flux measurements for Vega
and compiled a recommended SED from 3300--10500~\AA, including a revised
estimate of the monochromatic flux with a 25~\AA\ bandpass at 5556~\AA\ of
$3.44\times10^{-9}$~erg cm$^{-2}$ s$^{-1}$~\AA$^{-1}$ $\pm$1.5\% from an average
of five independent measures. See figure 6 of Bohlin et al. (2011, hereafter
B11) for a graphical comparison of the five F(5556) flux values with the H85 SED
for Vega. The ratio of this H85 SED to the STIS CALSPEC fluxes is illustrated in
Figure~\ref{hayes}. Any measured SED based on the H85 Vega fluxes, such as
BD+17$^{\circ}$4708 (Fukugita 1996, Bohlin \& Gilliland 2004b), differs from HST
based fluxes by more than 1\% over much of the wavelength range as in
Figure~\ref{hayes}.

Ten years after H85, Megessier (1995) reviewed the available absolute F(5556)
determinations and eliminated the low Oke \& Schild value because evidence
suggested that result was based on a faulty tungsten ribbon lamp. Another set of
results was scrapped because of discrepancies in the lab calibrations of their
tungsten lamps. A weighted average of the three remaining measurements yielded
$3.46\times10^{-9}$~erg cm$^{-2}$ s$^{-1}$~\AA$^{-1}$ $\pm$0.7\% for F(5556).
The rms scatter of 0.7\% is lower than the H85 1.5\%, because outliers were
eliminated. 

\subsubsection{Solar Analogs}

\begin{figure}
\centering
\includegraphics*[height=7in]{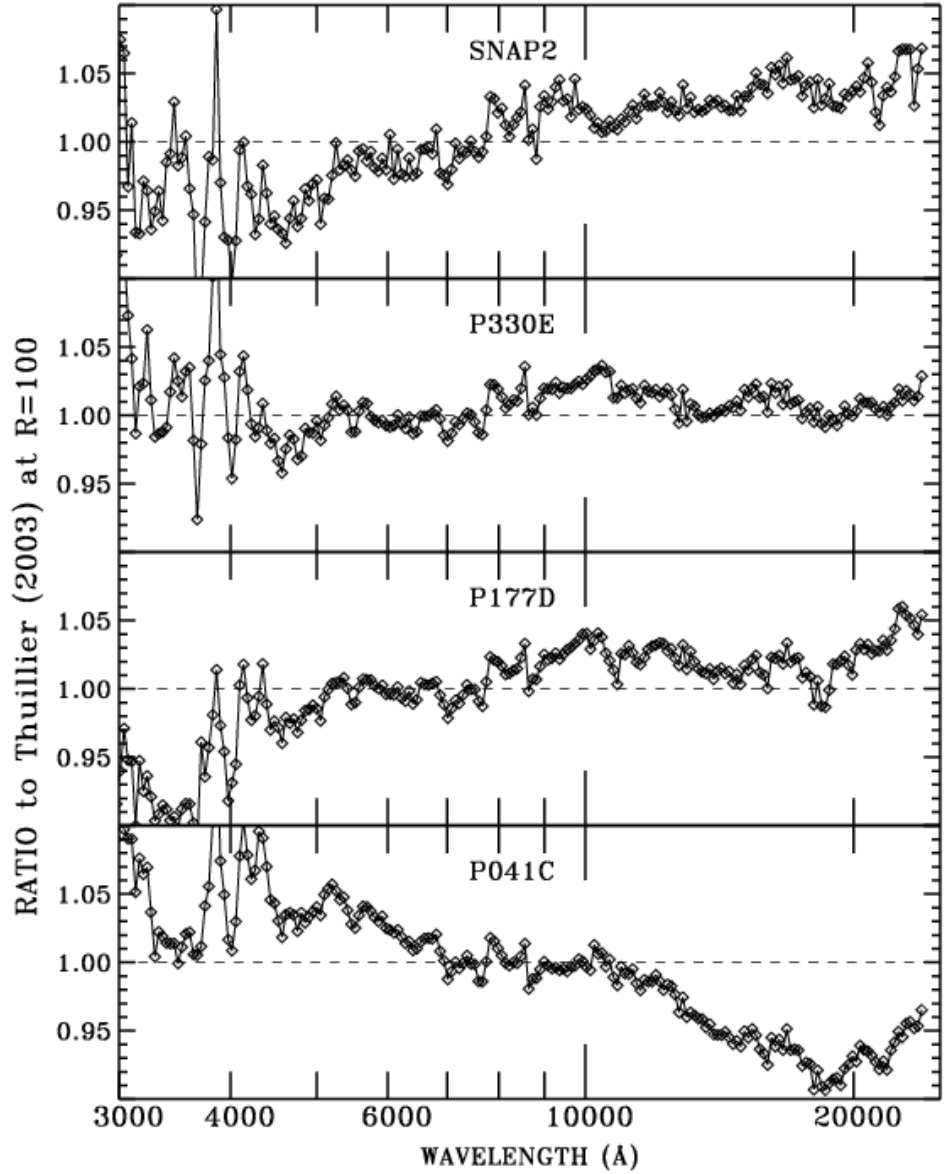}
\caption{\baselineskip=12pt
Ratio at R=100 of CALSPEC fluxes for solar type
stars to the Th03 solar SED, as normalized to unity at 7000--8000~\AA. 
\label{solar}} \end{figure}

The solar analog method of establishing primary stellar flux standards is to
assign the measured solar flux distribution to a star of the same G2V spectral
type, where the assumption is that the solar SED is known to better precision
than any star. One commonly adopted solar flux distribution is Thuillier et al.
(2003, Th03), where observations from the ATLAS and EURECA missions with the
Space Shuttle are referenced to the Heidelberg Observatory blackbody
absolute-flux standard. Figure~\ref{solar} compares existing CALSPEC flux
distributions for solar analogs (Bohlin 2010, hereafter B10) with the Th03 SED.
If  the solar twins, 18 Sco (G2V) and HD101364 (alias HIP56948, Melendez \&
Ramirez  2007) are ever observed with STIS in a photometric slit, those new data
should be compared to Figure~\ref{solar}.

While consistent to $\sim$10\% in Figure~\ref{solar}, these G stars do not agree
with each other or with the solar flux to 1\%. The de-reddening for the small
E(B-V) interstellar extinction values of B10 is not done but does not
significantly flatten or improve the consistency of the ratios of
Figure~\ref{solar}. Perhaps, a 1\% agreement with the Sun could be achieved with a
true solar twin.

At present, the solar analog technique has a lower precision than the HST
adopted WD technique, even if an unreddened star is a perfect match to the solar
line spectrum. The Th03 uncertainty at 2~$\mu$m is quoted as 1.3\% in table III
of Th03 but grows to "$\sim$2\%" in the conclusions section, which also quotes a
2--3\% uncertainty below 0.85~$\mu$m. The HST CALSPEC stellar SEDs have slightly
smaller formal uncertainties over the UV/optical/NIR wavelength range. In the
past, the brightness of the Sun facilitated accurate  measurements, especially
when bolometers were the primary detector. However with modern sensitive
detectors, comparison of laboratory flux standards with stars can be as precise
as for the Sun. In space, the solar brightness is actually a disadvantage,
because the large UV flux causes contamination of the optical components from
polymerized hydrocarbons.

\subsubsection{Modern}

Fundamental standards of irradience, i.e. flux, are maintained by NIST and are
traceable to a gold  melting point blackbody light source. Current NIST
2$\sigma$ uncertainties in the absolute responsivity of standard detectors are
0.2\% for Si photodiodes below 1~\micron\ (Brown et al. 2006) and 0.5\% for 
NIR photodiodes. Standard light sources with radiance temperatures as high as 
3000~K are available with an accuracy of 0.55\% at 5556~\AA\ 
(Fraser et al. 2007).

One currently funded program to establish primary stellar flux standards
relative to laboratory NIST irradiance standards is the ACCESS rocket program,
which will establish a few standards in the brightness range of Sirius to
V$\sim$9.5 (Kaiser et~al. 2007, 2010a, 2010b, 2012, 2013). The wavelength
coverage is 0.35--1.7~\micron\ with an accuracy goal of 1\% and a
spectral resolving power of R=500. Even though this wavelength range is
accessible from the ground, observations from above the atmosphere eliminates
that dominant source of uncertainty. ACCESS will be calibrated to both continuum
and emission line fundamental radiance standards. Using emission from tunable
lasers, the NIST SIRCUS facility (Brown et al. 2006) will provide an end-to-end
calibration transfer to ACCESS. These data also define the correction for
out-of-band spectral stray light using a matrix correction algorithm (Zong et
al. 2006, Smith et al. 2009). Stray light can cause serious errors when
measuring stellar spectral distributions which differ from the spectral
distribution of the calibrating light source. The use of a spectral light engine
will calibrate ACCESS using a continuum spectral energy distribution similar to
the spectral energy distribution of the various stellar targets (Brown et al.
2006, Smith et al. 2009).

Ground-based programs to measure stellar absolute fluxes seek to measure both
the instrumental response function and the optical transmission function of the
atmosphere (Stubbs \& Tonry 2012, Tonry et al. 2012). NIST-calibrated
photodiodes are utilized to measure the input to a telescope and establish the
total throughput quantum efficiency (QE). Empirical atmospheric transmission
is determined by water vapor as measured and matched to a MODTRAN
atmospheric transmission model to determine the real-time atmospheric
extinction.
The ground-based program, NIST STARS, uses lidar backscatter to measure the
atmospheric extinction at selected wavelengths and then fits a MODTRAN
model to determine the atmospheric extinction (McGraw et~al. 2010, Zimmer et~al.
2010). In principle, these techniques applied at high temporal cadence can
enable repeatable photometric stellar observation, even while atmospheric
clarity is unstable. A goal of NIST STARS is to measure the absolute flux of
stars across the sky with a 0.5\% precision.

\subsection{Infrared above 1~\micron}

Excellent reviews of the direct measurement of the physical fluxes of
stars in the infrared have been published (Rieke et al. 1985,
Price 2004, Rieke et al. 2008) and are summarized here.

 \subsubsection{Ground-based}

The direct measurement of stars in the infrared is complicated by the
atmospheric transmission that ranges from mostly transparent to totally opaque
over the 1--40~\micron\ wavelength range, where the water vapor and OH lines are
especially problematic. Ground-based observations necessarily
concentrate on measurements at wavelengths where the atmospheric opacity is low.

The earliest research measured absolute fluxes for a number of stars in
near-infrared bands, roughly corresponding to z, J, H, and K, by referencing to
blackbody sources at the telescopes (Walker 1969). This effort was extended to
mid-infrared wavelengths for a smaller sample of stars using Mars to transfer
the stellar measurements to laboratory blackbodies (Low et al. 1973, Becklin et
al. 1973, Rieke et al. 1985). Mainly because of the rapid variability of the
atmospheric transmission, the precision of ground-based IR photometry is limited
by the lack of accurate throughput measures of the instrumental bandpass that
relates photons above the atmosphere to detected quantum events as a function of
wavelength. Absolute spectrophotometry is always preferred to photometic average
fluxes over a filter bandpass; but to our knowledge, no IR spectrophotometry
beyond 1~\micron\ has been published with a pedigree that is directly based on
laboratory standards of absolute flux.

A significant body of work concentrated on precision measurements of Vega using
the same technique (Selby et al. 1980, 1983; Blackwell et al. 1983;
Mountain et al. 1985; Booth et al. 1989)
with later works providing growing evidence that the infrared measurements of
Vega deviate from extrapolations of optical fluxes into the IR. Detailed
observations and modeling have shown that Vega is a rapidly rotating star,
observed pole-on, and has a circumstellar disk that contributes to the stellar
flux starting around 2~\micron\ and extending to far-infrared wavelengths
(Aumann et al. 1984; Su et al. 2005, 2013; Aufdenberg et al. 2006; 
Sibthorpe et al. 2010; Defrere et al. 2011; Absil et al. 2013; Bohlin 2014).

\begin{figure}
\centering
\includegraphics*[height=5in]{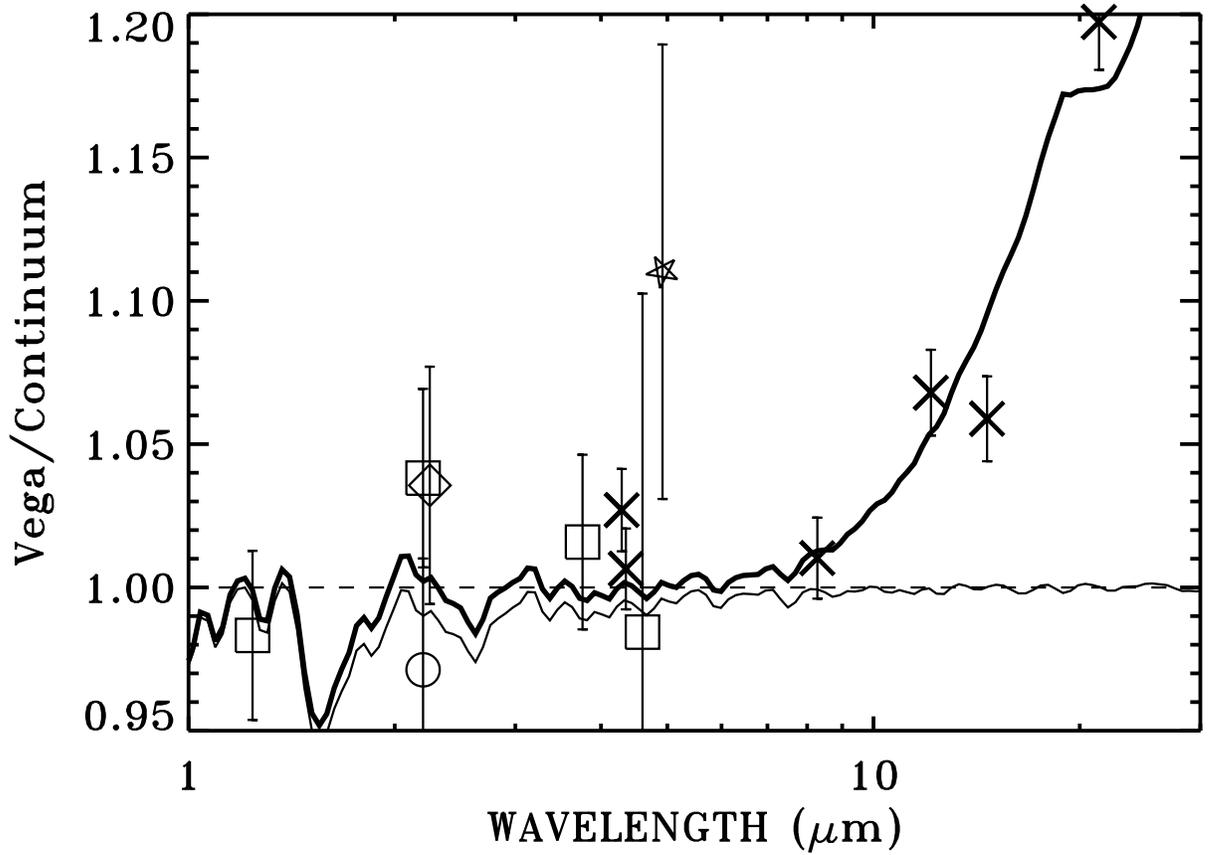}
\caption{\baselineskip=12pt
Comparison of historic IR absolute flux measures with the Vega SED of Bohlin
(2014) at a resolution of R=20 for the photospheric 9400~K model ($light~line$)
and the total flux including emission from the dust ring ($heavy~line$). All
values are divided by the theoretical photospheric continuum for
clarity of display. The various symbols for the physics-based
measurements are $circle$ (Selby et al. 1983),
$diamond$ (Booth et al. 1989), $star$ (Mountain et al. 1985), $squares$
(Blackwell et al. 1983), and X (MSX). \label{figir}} \end{figure}

However, comparisons with modern results suggest that the early flux
measurements are often correct within their quoted uncertainties. 
Figure~\ref{figir} shows the results of Bohlin (2014) for the Vega photosphere
and total flux of the dust plus photosphere in comparison to pioneering
ground-based measures of total absolute flux and MSX results (Price et al. 2004.
See next section.) For example, Booth et al. (1989) quote
$3.86\times10^{-11}~erg~cm^{-2}~s^{-1}~\AA^{-1}\pm4\%$ at 2.250~\micron\ with  a
bandpass of 10~\AA\ for Vega. The corresponding CALSPEC monochromatic flux from
the photosphere (i.e. alpha\_lyr\_stis\_007.fits) is
$3.676\times10^{-11}~erg~cm^{-2}~s^{-1}~\AA^{-1}$, while the contribution from
the dust in the K band is 1.26\% (Absil et al. 2013) for a total of
$3.72\times10^{-11}$~erg cm$^{-2}$ s$^{-1}$~\AA$^{-1}$. Thus, the Booth value is
only 3.7\% higher than the modern estimate and is within the Booth uncertainty
of 4\%. The Selby et al. (1983) measure of
$3.92\times10^{-11}~erg~cm^{-2}~s^{-1}~\AA^{-1}$ at 2.20~\micron\ is $\sim$3\%
lower than Bohlin (2014) but agrees within the quoted undertainty of 4\%.

\subsubsection{Airborne and Space-based}

The challenges associated with observing through the atmosphere have motivated
efforts to make direct measurements of stars in the infrared from airplanes and
from space-based telescopes. Witteborn et al. (1999) used the Kuiper Airborne
Observatory to provide direct measurements of a single star, $\alpha$~Boo, 
which demonstrates the challenges of measuring the absolute flux of stars using
observations through the atmosphere with the often complicated transfer to
laboratory blackbodies.

A space-based experiment with on-board calibration sources significantly
simplifies the direct measurement of stellar fluxes, which motivated the Spatial
Infrared Imaging Telescope (SPIRIT) III on the Midcourse Space Experiment (MSX
Mill et al. 1994). The MSX SPIRIT III observations covered the 4.3 to
21~\micron\ range in 6 bands and calibrated the observations of stars to five
emissive reference spheres that were ejected at various times during the 8
months of operations. The properties of these reference spheres were measured in
the laboratory, and their absolute infrared fluxes were determined as their
temperatures increased due to illumination by the Sun and Earth shine. The MSX
reference sphere calibration was checked by comparison to a set of predictions
for the fluxes of eight reference stars from the CWW network (Cohen et al.
1992a, Cohen 2007). The MSX measured mid-infrared fluxes have a quoted accuracy
of 1.4\% Price et al. (2004). Thus, the MSX mid-infrared measurements of the
Galactic plane and other selected areas provide a network of stars that have
been directly referenced to laboratory standards with fluxes calculated using
basic physics.

\subsection{Summary of Absolute Flux Zeropoint}

Bohlin (2014) observed the primary IR standard, Sirius, with STIS and normalized
an updated special Kurucz
model\footnote{http://kurucz.harvard.edu/stars/SIRIUS/} to the observed fluxes
at 6800--7700~\AA. The modeled IR extrapolation and the absolute Meggessier
visible flux are reconciled with the Midcourse Space Experiment (MSX) mid-IR
fluxes (Price et al. 2004) at 8--21\micron. In order to minimize both the
5556~\AA\ and MSX mid-IR residuals, the Megessier (1995) value with its 0.7\%
quoted uncertainty must be multiplied by 0.9945 to achieve a properly weighted
average. The largest residual is for the MSX A band at 8.3~\micron, where the
residual is 1.8\%, i.e. 1.3$\sigma$. The final result from Bohlin (2014) is
F(5556)=$3.44\times10^{-9}$~erg cm$^{-2}$ s$^{-1}$~\AA$^{-1}$ with a formal
uncertainty of 0.5\%. While NLTE models establish the relative SEDs of the three
primary HST standards as a function of wavelength, their overall absolute flux
level is set by F(5556) for Vega, as discussed below.

\section{Using Models to Establish SEDs}

\subsection{History and Rationale}

Despite valiant efforts to tie stars to lab flux standards in the 1970s,
inconsistencies in the available standard stars still existed, as illustrated in
some of the above figures. Thus, D. Finley and J. Holberg (Finley et al. 1984,
Holberg et al. 1986, Finley et al. 1990) suggested the use of pure hydrogen WD
model atmospheres for the UV flux calibration of the IUE satellite (Bohlin et al. 1990).
Pure hydrogen WDs are preferred, because model atmosphere calculations are
greatly simplified with only one element to consider. As the basis for all HST
absolute fluxes, Bohlin, Colina, \& Finley (1995) adopted the D. Koester model
atmosphere SEDs calculated in local thermodynamic equilibrium (LTE) for G191B2B,
GD153, GD71, and HZ43. Subsequently, HZ43 fell off this list of primary flux
standards because of an M star companion that contaminates the STIS observations
in the visible and IR (Bohlin et al. 2001). For the remaining three stars,
interstellar reddening from the dust reduces the flux by $<$0.6\% longward of
1150~\AA, given the strict limits on $E(B-V)$ derived from the low hydrogen
column densities and the galactic average $N(\mathrm{HI})/E(B-V)=4.8\times10^{21}$ of
Bohlin et al. (1978). Of our three stars, G191B2B has the largest hydrogen
column density $N(HI)=2.2\times10^{18}$ (Rauch et al. 2013, hereafter RWBK),
which corresponds to $E(B-V)=0.0005$. Dupuis et al. (1995) find
$N(\mathrm{HI})<1\times10^{18}$ for GD153 and GD71. For G191B2B, the small reddening has
been applied to the model according to the  $R(V) = 3.1$ reddening curve of
Cardelli et al. (1989) down to 2000~\AA, where the LMC  curve of Koornneef and
Code (1981) for small grains is substituted, because the steeper far-UV slope
fits the data better.


In order to establish the effective temperature and surface gravity
($T_\mathrm{eff}$ and $\log g$) for the pure hydrogen WDs, the calculated model
lines are fit to the observed Balmer line profiles (e.g. Finley et~al. 1997,
hereafter FKB). FKB used Koester LTE model line profiles to fit the Balmer
lines. However, non-LTE (NLTE) calculations should be a better representation of
the actual stellar physics. Consequently, Bohlin (2003) adopted the Hubeny
Tlusty NLTE models (version 203) for pure hydrogen atmospheres (Hubeny \& Lanz
1995). However, the $T_\mathrm{eff}$ and $\log g$ values originally used for the
NLTE models were those derived from the LTE fits. Recently, this deficiency has
been rectified by Gianninas et al. (2011, G11), who fit new Balmer line
observations of the WDs with updated Tlusty NLTE models that
include improved Stark broadening of the Balmer lines (Tremblay \& Bergeron
2009).

\subsection{The HST Absolute Fluxes}

\subsubsection{The WD Models}

The absolute flux at 5556~\AA\ (5557.5~\AA\ in vacuum) of the NLTE models for
the three primary WD standards is set by the STIS spectrophotometry of Vega
(Bohlin \& Gilliland 2004a, Bohlin 2007) relative to the WDs and by the absolute
monochromatic flux at 5556~\AA\ for Vega of F(5556)=$3.44\times10^{-9}$~erg
cm$^{-2}$ s$^{-1}$~\AA$^{-1}$ as discussed above. The small uncertainty of 0.5\%
in this F(5556) value affects the overall level and not the shape (i.e.
``color'') of the WD models used for \emph{HST} flux calibrations. Despite
suggestions that Vega is a variable star, H85 discusses the evidence and
concludes that any variability "must be less than 0.01 mag". Engelke et al.
(2010) present evidence for a 0.08 mag variation of Vega at visible wavelengths;
but Bohlin (2014) demonstrated that this apparent variability seen in Hipparcos
data is actually just a symptom of pulse counting saturation for this bright
star.

Once the three WD flux distributions are fully defined by their NLTE models as
normalized to the reconciled absolute visible/IR level, the STIS flux
calibration proceeds as outlined in Section 2. The fully corrected $N_e$ count
rate in electrons~s$^{-1}$ as a function of wavelength for each STIS observation
of the three primaries is extracted from the STIS images using an extraction
height of 11 pixels for the UV MAMA data and 7 pixels for the CCD images. The
procedure for correction to infinite extraction height is outlined in Bohlin
(1998). Each $N_e$ spectrum is matched to the standard star flux, $F_{\lambda}$,
with wavelengths adjusted to the instrumental rest frame and with the model
smoothed to the STIS resolution. Because the STIS LSF is not precisely known and
the model line profiles are not perfect, the absorption line regions are masked
before the sensitivities $S_\lambda=F_{\lambda}/N_e$ per Equation (\ref{fspl})
are fit with splines. The $IDL$ procedure $splinefit$ is used for fitting the
1024 pixel sensitivities with 50--60 nodes per each of the five STIS low
dispersion modes. The spline fits for all three WDs are averaged with equal
weight for each star to get the STIS sensitivities that are used to calibrate
other STIS observations and establish secondary flux standards. Because G191B2B
is too bright for the two UV MAMA modes, those sensitivity functions are defined
only by GD71 and GD153. All STIS observations of flux standards utilize the wide
52X2 arcsec slit to avoid variable slit losses.

\begin{figure}
\centering
\includegraphics*[height=7.5in]{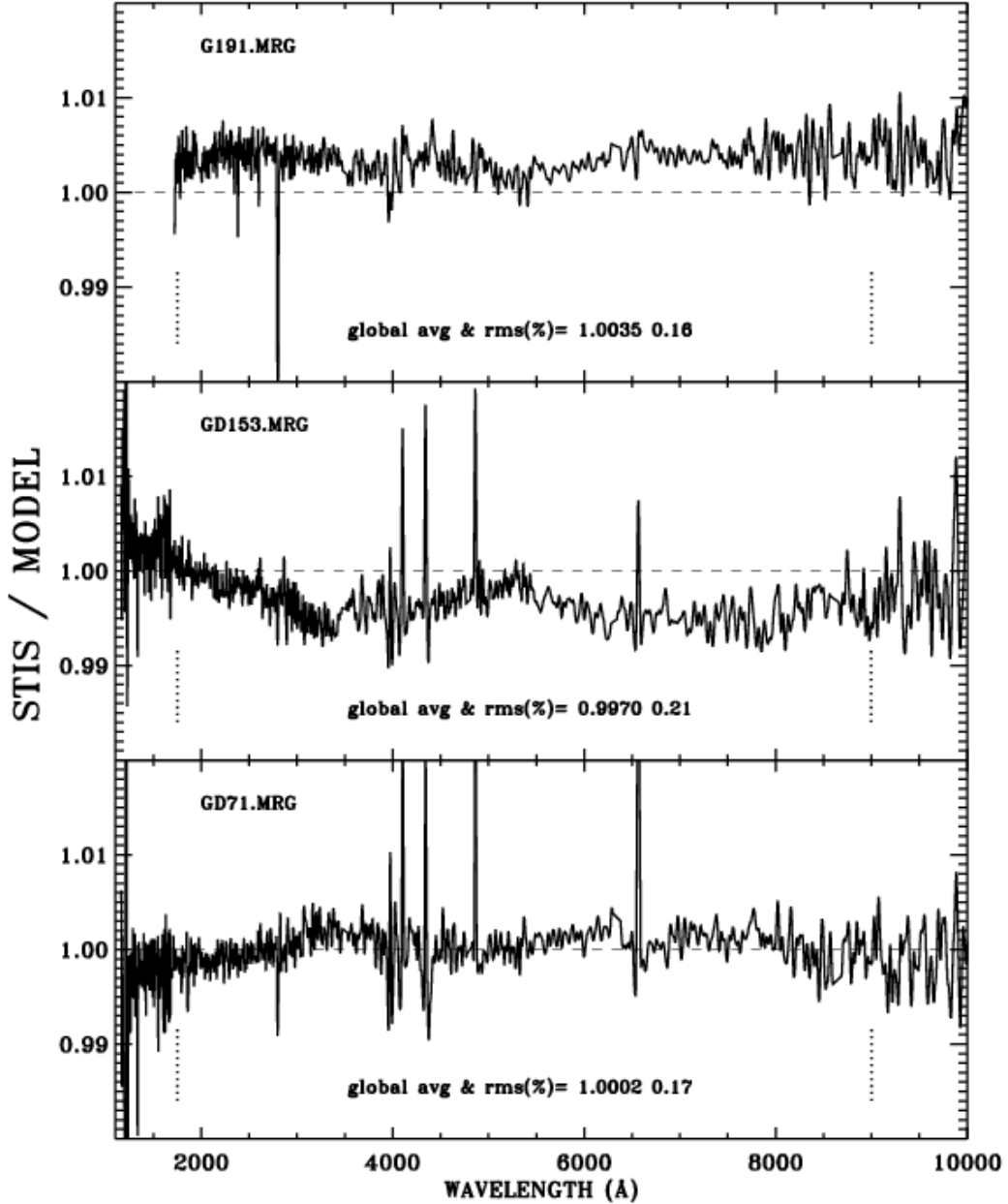}
\caption{\baselineskip=12pt
Ratio of old STIS flux distributions to the old Tlusty 203 NLTE 
models for pure hydrogen that were used to define the old fluxes.
The average ratios and rms scatter between the
vertical dotted lines at 1750 and 9000~\AA\ are written on the
plots. Narrow
band differences are evident in the Balmer lines. G191B2B has a strong MgII
interstellar or circumstellar absorption feature at 2800~\AA. There are no data
for G191B2B below 1700~\AA, because the star is too bright for the STIS MAMA
detectors.
\label{oldcal}} \end{figure}

Figure~\ref{oldcal} illustrates the internal residuals after the old STIS
low-dispersion fluxes of the WDs are divided by the Tlusty 203 pure hydrogen
NLTE models used to define the old flux calibration. The agreement at the
sub-percent level demonstrates that the calibrated fluxes agree with their
reference SEDs to better than $\sim$0.5\% over the 1150--10000~\AA\ wavelength
range, except for a few narrow bands at absorption lines. Thus, any updates to
the modeled SEDs of these primary standard should retain the same sub-percent
level of internal agreement. 

\begin{figure}
\centering
\includegraphics*[height=7.5in]{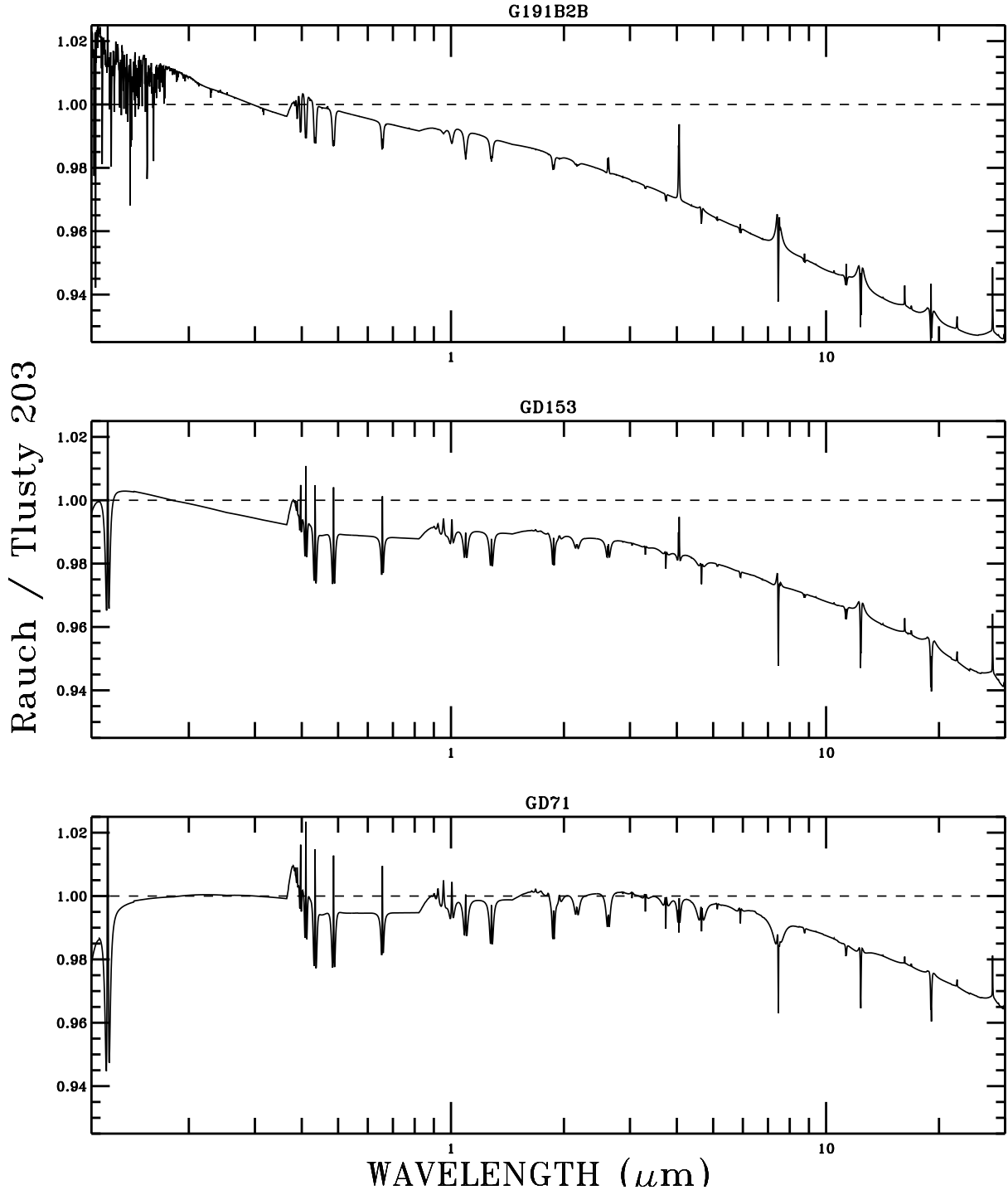}
\caption{\baselineskip=12pt
Ratio at R=500 of the new Rauch model fluxes to the pure hydrogen Tlusty 203 models 
that previously defined the three primary WD SEDs. All models are computed in
NLTE with the parameters in Table 1. For GD153 and GD71, the new models are for pure
hydrogen, while RWBK have computed a full metal line blanketed model of G191B2B
that matches high dispersion observations of the UV absorption lines. From
Ly$\alpha$ to 1~\micron\ over the STIS wavelength range, the three ratio plots
are the same to $\sim$2\% in the continuum. Narrow band differences are evident
in the hydrogen line profiles. The Rauch models include the new 0.6\% gray
reduction of the fluxes  with respect to the old Tlusty normalization (Bohlin
2014). \label{newsed}}\end{figure}

\begin{figure}
\centering
\includegraphics*[height=7.5in]{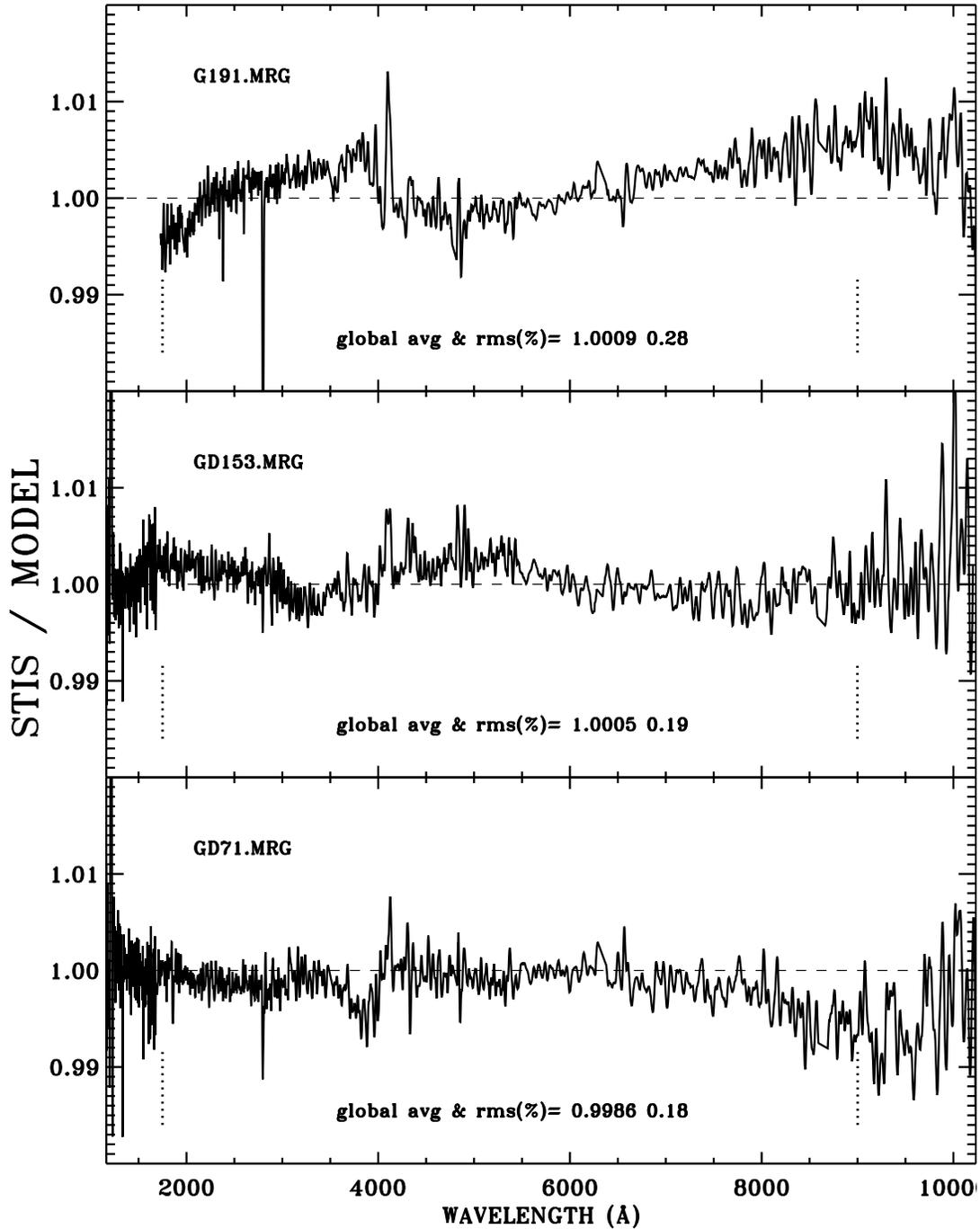}
\caption{\baselineskip=12pt
Ratio of new STIS spectrophotometry as in Figure~\ref{oldcal} to the new primary
standards made from the Rauch models in Figure~\ref{newsed}. Notice the smaller
residuals in the Balmer lines for GD153 and GD71 in comparison with
Figure~\ref{oldcal}. \label{newcal}}\end{figure}

\begin{deluxetable}{cccccccc}
\tablewidth{0pt}
\tablecolumns{8}
\tablecaption{The Primary WD Stars}
\tablehead{
\colhead{Star} &\colhead{V\tablenotemark{a}} &\colhead{Sp. T.} 
&\colhead{FKB $T_\mathrm{eff}$} &\colhead{FKB $\log g$} 
&\colhead{G11 $T_\mathrm{eff}$} &\colhead{G11 $\log g$}
&\colhead{Unc.$T_\mathrm{eff}$}}
\startdata
G191B2B &11.781 &DA.8  &61193 &7.492 &60920\tablenotemark{b} &7.55 &993 \\
GD153   &13.346 &DA1.2 &38686 &7.662 &40320 &7.93 &626 \\
GD71    &13.032 &DA1.5 &32747 &7.683 &33590 &7.93 &483 \\
\enddata

\tablenotetext{a}{G191B2B--Landolt and Uomoto (2007), GD153--Landolt (1995
 private comm.), GD71--Landolt (1992)}

\tablenotetext{b}{The $T_\mathrm{eff}$ and gravity of the best fitting metal 
line-blanketed model are 59000~K and $\log g=7.6$.}

\end{deluxetable}

Figure~\ref{newsed} illustrates the difference between the original Tlusty 203 
and the new Rauch NLTE models with $T_\mathrm{eff}$ and $\log g$ from G11 for
GD153 and GD71. Figure~\ref{newsed} shows that the Rauch models from the
registered Virtual Observatory service  TheoSSA\footnote{Theoretical Stellar
Spectra Access, \url{http://dc.g-vo.org/theossa}} (Werner et al. 2003) change by
similar amounts over 1150--10000~\AA\ for each star relative to the old Tlusty
SEDs. The pure hydrogen model for G191B2B does {\em not} show the same ratio as
the two cooler stars, but that model is inappropriate because of the trace metal
lines observed in the UV. Instead, a special line blanketed NLTE model of RWBK
is compared with the Hubeny SED for G191B2B in Figure~\ref{newsed}. While RWBK
found $T_\mathrm{eff}=60000 \pm 2000$~K, a $T_\mathrm{eff}=59000$~K model is
within the uncertainty and is more consistent with the relative UV flux of the
three stars. The parameters defining the three primary flux standards appear in
Table 1. After a complete re-calibration (see section 2) using the new set of
model SEDs, Figure~\ref{newcal} illustrates the new internal STIS residuals,
which are comparable to the old residuals in Figure~\ref{oldcal}. The net
changes in STIS fluxes due to this update of the three prime standards are
$\lesssim$~1\%. Updated fluxes that are based on the new Rauch model SEDs are
identified in the CALSPEC database by delivery dates in 2013 or later.

\subsubsection{The Uncertainties}

The uncertainty in the overall level of the absolute fluxes is the 0.5\% for the
F(5556)=$3.44\times10^{-9}$~erg cm$^{-2}$ s$^{-1}$~\AA$^{-1}$ flux of Vega. The
possible error in the WD SEDs could be larger, if Vega is variable or if STIS is
non-linear over the large dynamic range of 12--13 mag for the measured ratios of
the WDs to Vega at 5556~\AA. While the variability of Vega remains
controversial, there is no charge transfer efficiency (CTE) loss in the STIS CCD
observations for such a bright star according to the CTE correction formula of
Goudfrooij et al. (2006).

\begin{figure}
\centering
\includegraphics*[height=7.5in,trim=0 0 0 0]{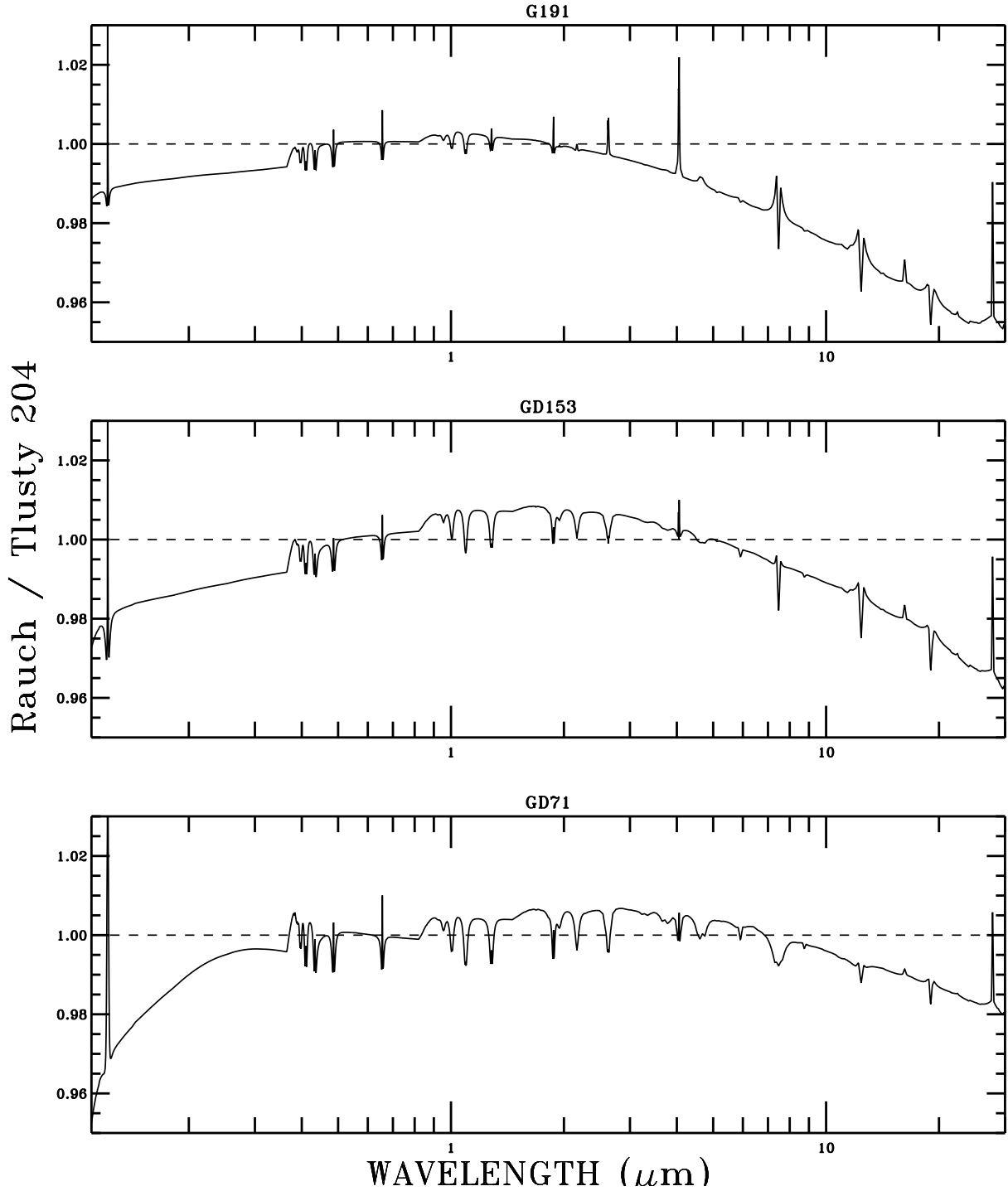}
\caption{\baselineskip=12pt
Ratios of pure hydrogen models for the results from the Tlusty 204 and Rauch NLTE
codes, both using the Stark broadening profiles of Tremblay \& Bergeron 2009. 
The resolution is R=500 and both models for each star are normalized to
the same value at 5566~\AA. The deviation from unity for the three stars provides a measure of the
uncertainty of using WD models to represent the SEDs of actual stars; and
the worst agreement at the shorter wavelengths is for the coolest star GD71.
\label{moderr}}\end{figure}

More important is the uncertainty in the slopes of the adopted SEDs, i.e. the
ratio of the model fluxes to their flux at 5556~\AA. One measure of this
uncertainty in the new set of primary standard models is the difference between
sets of pure hydrogen NLTE models with the same $T_\mathrm{eff}$ and $\log g$.
Our Tlusty NLTE models (version 204) include the  Tremblay \& Bergeron (2009)
Stark profiles (see also Tremblay et al. 2011, G11),  and incorporate similar
physics to the Rauch models. The model ratios appear in Figure~\ref{moderr} for
the G11 determinations of $T_\mathrm{eff}$ and $\log g$. In the STIS range, the
Tlusty 204 models are systematically higher by up to 3\% in the far-UV at
Ly-$\alpha$ for GD71, while the IR fluxes are also higher, with a worst case
difference of nearly 5\% at 30~\micron\ for G191B2B. The pairs of models agree
to $\sim$1\% from 0.2--5~\micron.

To understand the differences between the Rauch (TMAP) and Tlusty NLTE results,
the actual computer codes should be compared in detail. However in the IR, the
LTE vs. NLTE differences (see B11) are significantly larger than the
differences between the NLTE codes, which implies that IR fluxes at the few 
percent level are rather sensitive to the input microphysics and to the NLTE
stratification of the upper layers of the model atmospheres. 


\begin{figure}  
\centering  
\includegraphics*[height=4.51in,trim=70 0 0 0]{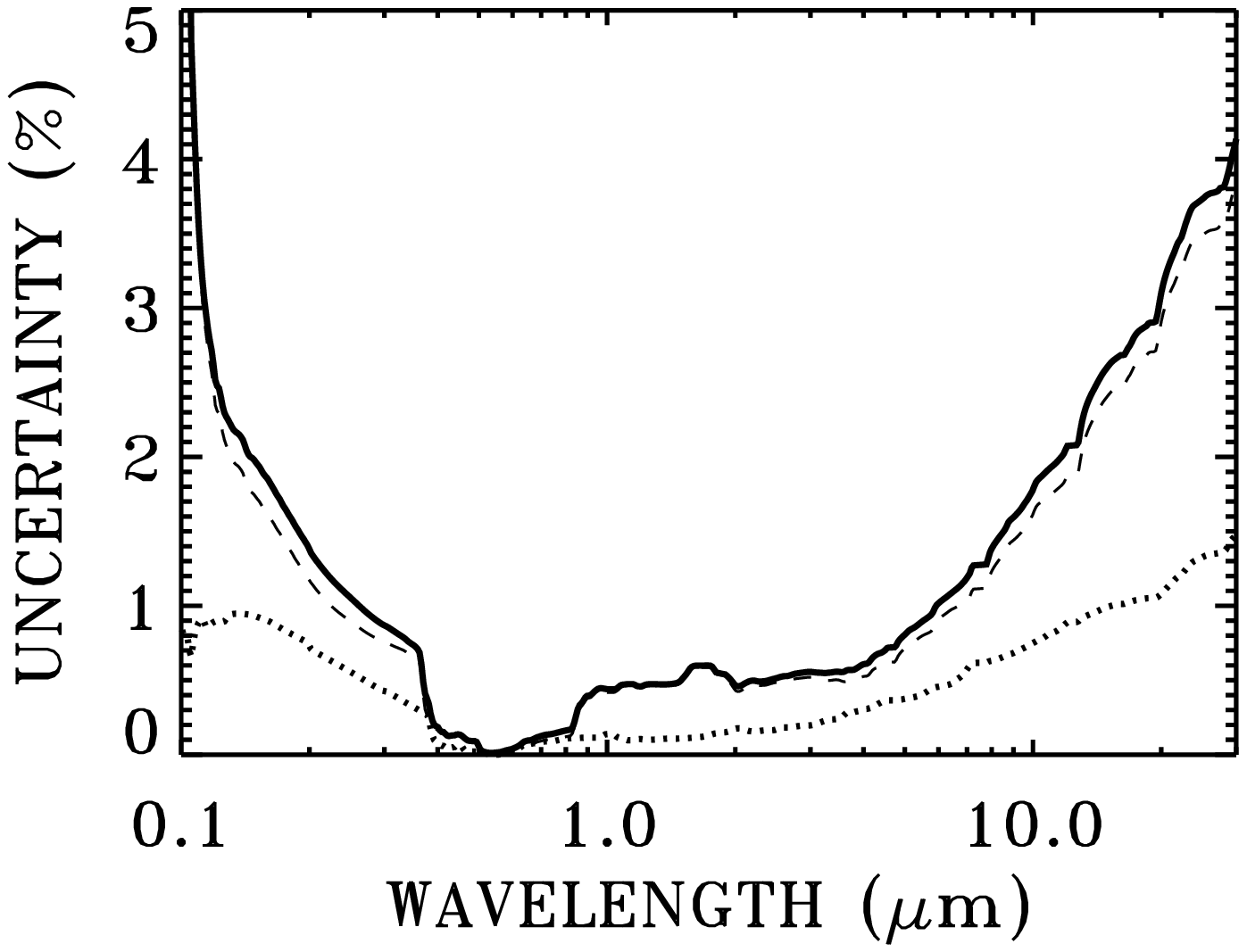}
\caption{\baselineskip=12pt
$Dotted~line$ -- rms uncertainty in the WD flux scale from formal errors in
$T_\mathrm{eff}$ from G11. $Dashed~line$ -- The dominant rms uncertainty, as
calculated from differences between pure hydrogen Tlusty 204 and Rauch 
NLTE models
with the same $T_\mathrm{eff}$ and $\log g$.  $Heavy~solid~line$ -- Combination
in quadrature of the above two uncertainties. These curves are all relative to
the flux at 5556~\AA, where the relative uncertainty is zero by definition.
\label{uncert}}\end{figure}

Another smaller contributor to the uncertainty in the model slope is caused by
the $T_\mathrm{eff}$ measurement errors from G11; but the uncertainties of 0.05
in $\log g$ cause, for example, at most a 0.2\% continuum flux difference for
G191B2B and are neglected. The dotted line in Figure~\ref{uncert} represents the
rms differences from unity for the ratios of nominal temperature models to
models that differ by the uncertainty in degrees Kelvin from the last column of
Table 1. The dashed line in Figure~\ref{uncert} is the rms average difference
from unity for the three ratios of Figure~\ref{moderr} and represents the
uncertainty in the modeling procedure, while the heavy solid line is the
combined estimate of total uncertainty in  the WD flux system. Because the model
SEDs are all normalized to unity at 5556~\AA, the uncertainties are all relative
to 5556~\AA, where the uncertainty in flux due to slope errors is defined to be
zero. When the primary stars are used for an instrumental calibration, the
strong hydrogen lines are masked to avoid errors due to imprecisely known
instrumental line-spread-functions (LSF). Also, the modeling precision in the
line cores and wings is less than the precision in the continuum. Thus, the
effects of the lines are also removed in Figure~\ref{uncert}.

While Figure~\ref{uncert} suffices for simple minded estimates of the ensemble
systematic uncertainty of the HST flux system, correlations of the errors over
the wavelength range require the more comprehensive covariance matrix to fully
characterize the  uncertainties (Jeach 1985). This $573\times573$ matrix is
available from the CALSPEC archive as the binary table WDcovar.fits in units of
fractional error. The 573 wavelength bins represent a resolution of R=100 from
1000~\AA\ to 30\micron, while the square of the heavy solid line in
Figure~\ref{uncert} is the diagonal of the covariance matrix after dividing the
values from the Figure by 100 to get the fractional error.

A set of WD models for the primary standards provides complete wavelength
coverage for a system of absolute fluxes and for the zero points of the various
standard magnitude systems; see for example Holberg \& Bergeron (2006) and
Pickles (2010). However, differences reach 2\% at 8~\micron\ in
Figure~\ref{moderr} between the independent G191B2B models, which suggests 
that, pending further investigation, these WD SEDs are not appropriate James 
Webb Space Telescope (JWST) standards out to 30~\micron. 

In summary, the likely systematic errors in the slope of our WD based flux
system is the heavy solid line in Figure~\ref{uncert} or the covariance matrix
WDcovar.fits. In order to get an estimate of the total systematic uncertainty at
any wavelength, the uncertainty relative to 5556~\AA\ must be combined with the
uncertainty in F(5556) for Vega, which is 0.5\%.
Since the covariance is a measure of the uncertainty $\sigma^2_{ij}$, a value of
$0.005^2=0.000025$ must be added to every matrix element to convert the
covariance from values relative to 5556~\AA\ to total $\sigma^2$. Our ensemble
uncertainty estimates are applicable to instrumental flux measures when
observations with equal weight for all three primary WDs are utilized for the
flux calibration. For the best transfer of the WD flux system to an instrumental
calibration, the observational data set for the three primary WDs should be
robust enough to make statistical errors negligible, while spacing of the
observations over the instrumental lifetime serves to reduce errors in
accounting for any time dependent effects.

Of course, the total uncertainty in the observed flux for a sparsely observed
program star is the uncertainty in the particular observational data set
combined with the ensemble systematic uncertainty. For example, the
signal-to-noise, the broadband repeatability, and any non-linearities contribute
to larger errors in the flux of any program star. See figure 1 of Bohlin \&
Gilliland (2004b) for a discussion of statistical uncertainty for STIS.

The new models for the HST WD standard stars helps explain part of the problem
with the WD fluxes in comparison with the Spitzer fluxes (B11). The worst
discrepancy found by B11 was a 4$\sigma$ difference of 12\% for G191B2B with
IRAC4 at 8~\micron. The 4\% lower flux at 8~\micron\ for the new G191B2B SED in
Figure~\ref{newsed}, reduces the discrepancy to 8\% with less than a 3$\sigma$
significance.

\subsection{IR Fluxes of Normal Stars}

\begin{deluxetable}{lcclrrrrrcc}
\tabletypesize{\scriptsize}
\tablewidth{0pt}
\tablecolumns{17}
\tablecaption{Secondary Standards with IR Fluxes Defined by CK04 Models\tablenotemark{a}}
\tablehead{
\colhead{Star} &\colhead{$R.A.$} &\colhead{$Decl.$} &\colhead{$Sp.T.$} &\colhead{$V$}
&&\colhead{$T_\mathrm{eff}$} &\colhead{$\log g$} &\colhead{$[M/H]$} &\colhead{$E(B\!-\!V)$}
&\colhead{$\chi^2$}\\
&\multicolumn{1}{c}{J2000} &\multicolumn{0}{c}{J2000}
}

\startdata
$\xi^{2}$ Cet  &02 28 09.54 &+08 27 36.2  &B9III &4.30 &&10360 &3.90 &$-0.61$ &0.001 &1.35  \\
$\lambda$ Lep &05 19 34.52 &$-$13 10 36.4 &B0.5V &4.29 &&27920 &4.30 &0.10 &0.017 &5.58  \\
$\mu$ Col     &05 45 59.89 &$-$32 18 23.2 &O9.5V &5.18 &&29840 &4.00 &0.12 &0.003 &7.00  \\
10 Lac       &22 39 15.68 &+39 03 01.0     &O9V   &4.88 &&29880 &3.80 &0.03 &0.069 &7.67  \\
\\
HD014943   &02 22 54.68 &$-51$ 05 31.7 &A5V & 5.91 &&8040 &3.90 &$-0.10$ &0.037 & 3.05   \\
HD37725    &05 41 54.37 &+29 17 50.9 &A3V & 8.35  &&8140 &4.15 &$-0.32$ &0.024 & 1.93   \\
HD116405   &13 22 45.12 &+44 42 53.9 &A0V & 8.34  &&10800&4.10 &$-0.37$ &0.002 & 0.76   \\
BD+60 1753 &17 24 52.27 &+60 25 50.7 &A1V & 9.67  &&9440 &4.00 &$-0.12$ &0.023 & 0.49   \\
HD158485   &17 26 04.84 &+58 39 06.8 &A4V & 6.50  &&8700 &4.30 &$-0.38$ &0.066 & 2.56   \\
1732526    &17 32 52.64 &+71 04 43.2 &A3V &12.53  &&8860 &4.10 &$-0.20$ &0.061 & 2.64   \\
1743045    &17 43 04.48 &+66 55 01.6 &A5V &13.6   &&7350 &3.50 &-$0.47$ &0.014 & 0.71   \\
HD163466   &17 52 25.37 &+60 23 46.9 &A2  & 6.86  &&7880 &3.60 &$-0.56$ &0.030 & 4.77   \\
1757132    &17 57 13.24 &+67 03 40.8 &A3V &12.01  &&7860 &4.10 &\phn~0.34 &0.071 & 0.88  \\
1802271    &18 02 27.17 &+60 43 35.7 &A3V &11.98  &&9070 &4.10 &$-0.58$ &0.024 & 0.80   \\
1805292    &18 05 29.28 &+64 27 52.0 &A1V &12.28  &&8540 &4.00 &$-0.13$ &0.033 & 0.50   \\
1808347    &18 08 34.70 &+69 27 28.7 &A3V &11.69  &&7900 &3.90 &$-0.73$ &0.027 & 1.32   \\
1812095    &18 12  9.60 &+63 29 42.2 &A5V &12.01  &&7750 &3.60 &\phn~0.06 &0.002 & 1.34  \\
HD180609   &19 12 47.20 &+64 10 37.2 &A0V & 9.41  &&8560 &4.00 &$-0.55$ &0.042 & 0.65   \\
\\
C26202     & 3 32 32.84 &$-27$ 51 48.6 &F8IV  &16.64 &&6200 &4.40 &$-0.52$ &0.053 &0.67 \\
HD037962   &05 40 51.97 &$-31$ 21 04.0 &G2V & 7.85   &&6000 &5.00 &\phn~0.00 &0.059 &1.87 \\
HD038949   &05 48 20.06 &$-24$ 27 49.9 &G1V & 8.0    &&6080 &4.20 &$-0.12$ &0.016 &1.11	\\
HD106252   &12 13 29.51 &+10 02 29.9 &G0  & 7.36    &&5940 &4.70 &\phn~0.00 &0.016 &1.40	\\
P041C\tablenotemark{b} &14 51 57.98 &+71 43 17.4 &G0V &12.16 &&6020 &4.15 &\phn~0.07 &0.034 &0.79 \\
P177D      &15 59 13.57 &+47 36 41.9 &G0V  &13.36   &&5880 &3.80 &$-0.11$ &0.052 &0.98 \\
SF1615+001A &16 18 14.23 &+00 00 08.6 &G0-5  &16.75 &&5880 &4.30 &$-0.73$ &0.118 &0.45 \\
SNAP-2     &16 19 46.11 &+55 34 17.8 &G0-5  &16.23  &&5760 &4.90 &$-0.36$ &0.034 &0.74 \\
P330E      &16 31 33.82 &+30 08 46.5 &G2V  &12.92   &&5920 &4.80 &$-0.13$ &0.051 &1.59 \\
HD159222   &17 32 00.99 &+34 16 16.1 &G1V & 6.56    &&5780 &3.90 &\phn~0.00 &0.001 &1.80 \\
HD205905   &21 39 10.15 &$-27$ 18 23.7 &G2V & 6.74  &&5920 &4.10 &\phn~0.00 &0.025 & 1.91 \\
HD209458   &22 03 10.77 &+18 53 03.5 &G0V  &\phn7.63 &&6100 &4.20 &$-0.04$ &0.003 &0.53 \\
\enddata

\tablenotetext{a}{The following stars with data are omitted: HD27835 and
HD60753-double stars, HD165459-dust ring, 1739431-bad focus, 1740346-dust ring,
1812524-poor fit to model atmosphere.}

\tablenotetext{b}{P041C has an M companion 0.57arcsec away Gilliland 
\& Rajan (2011)}

\end{deluxetable}

Because of the uncertainty in the IR fluxes of the WDs and because of the
sparsity of bright pure hydrogen WDs, an alternate method is required for
establishing a network of standard stars with known IR absolute fluxes. Longward
of 1~\micron, an extensive network of standard stars based on A-star models of
Vega and Sirius has been established in a sequence of papers I--XIV (e.g. Cohen
et al. 1992b, Cohen et al. 2003). The essense of this technique is to find a
model from a published
grid\footnote{http://wwwuser.oats.inaf.it/castelli/grids.html},  e.g. Castelli
\& Kurucz (2004, hereafter CK04) that fits the observed shorter wavelength
fluxes so that the longer wavelength part of the model establishes the absolute
IR fluxes of the standard star. This technique was used by Bohlin \& Cohen
(2008) for A stars and by B10 for G stars with a fitting technique in four
parameters, $T_\mathrm{eff}$, $\log g$, the metallicity [M/H], and the reddening
E(B-V). With the switch to Rauch NLTE models to define the primary WD SEDs, the
measured HST fluxes change slightly; and new fits are required. Table 2 includes
the revised best fit parameters for the Bohlin \& Cohen and B10 stars that now
have both STIS and NICMOS spectrophotometry. Additional new O, B, A, and G
standards have only STIS spectrophotometry. The fitting technique is modified to
use chi-squared minimization rather than the previous minimization of the $rms$
residuals. These stars are prime candidate JWST flux standards; and their SEDs
consisting of STIS fluxes plus the modeled IR extensions are in the CALSPEC
database.

Despite changes in the SEDs of the three prime reference standards, improved
observational data for the model fitting, and improved fitting techniques, the
largest change in the modeled IR fluxes of Bohlin \& Cohen or B10 is 3\% at
30~\micron\ for 1812095, which corresponds to a change in $T_\mathrm{eff}$ from
8250~K to 7750~K. Because the IR fluxes for G to A stars approach a
Rayleigh-Jeans distribution, the fitted fluxes are robust and not a strong
function of temperature. The reason for such large changes for 1812095 is that
Bohlin \& Cohen fitted only to the NICMOS spectrophotometry, which has a short
wavelength cutoff of 0.8~\micron, while the new fit is constrained by new STIS
data down to 0.114~\micron. For the G stars of B10, the original IR flux
extrapolations differ typically by $<$1\% from the revised parametizations in
Table 2, because STIS data already existed for the B10 analysis. 

Just as for the three primary WDs, the systematic errors of calculated model
grids likely dominate the uncertainty in the extrapolated IR fluxes. For the G
stars, B10 compared results for MARCS models (Gustafsson et al. 2008) with the
CK04 grid and found models from both grids that agree with the STIS
fluxes to $\sim$0.5\% at shorter wavelengths. But the MARCS grid does not extend
past 20~\micron, while the CK04 grid has rather coarse wavelength spacing with a
total of just 1221 sample points and only one point between 10 and 40~$\mu$m. 
Recently, M\'esz\'aros et al. (2012) have computed  an expanded set of the
Kurucz ATLAS9 models with updated abundances. Unfortunately, the wavelength grid
is even sparser with only 333 points. A third independent model grid, an update
of the CK04 models with good wavelength resolution, and an extension of the
MARCS grid beyond its current 20~\micron\ limit are required to establish
confidence in the IR flux distributions. Shortward of 20~\micron, the best-fit
CK04 models agree with the best-fit MARCS models to 1\% for the G stars in broad
continuum bands. Combining this 1\% with a 1\% systematic uncertainty in the
\emph{HST} WD flux scale at the anchor point of the fits at 1--2~\micron\ from
Figure~\ref{uncert} results in an estimate of a possible systematic 2\%
uncertainty of the broadband IR fluxes of individual G stars near 20~\micron\
with respect to the V band. At 20~\micron, this 2\% uncertainty is less than the
3\% uncertainty for WDs from Figure~\ref{uncert}.

For the A stars, a second independent model grid is needed, because the MARCS
models are limited to maximum temperatures of 8000~K. Both the original Cohen
SEDs and the newer Bohlin \& Cohen (2008) SEDs are extrapolations of the
measured fluxes into the IR using models that are based on computer code which
is traceable to R. Kurucz. The best check on these IR fluxes of the A stars was
from the two models provided by T. Lanz at 9400~K and 8020~K. While the
agreement of the Lanz SED with CK04 is within 2\% for the 9400~K model, the Lanz
SED is brighter than CK04 by 4\% at 10~\micron\ for the 8020~K model, which
raises the question of the accuracy of model SEDs, in general. A new high
resolution grid is currenty being produced by  Sz. M\'esz\'aros and should
significantly improve the situation.

Cooler stars, such as K type, are often considered for extrapolating model SEDs
to longer wavelengths. However, the extensive contribution of molecular line
blanketing complicates the models, which suggests that hotter stars should be
superior IR flux standards whenever the SED depends on model calculations.

\subsection{Non-WD Models as Standard Star SEDs}

\begin{figure} 
\centering 
\includegraphics*[height=8in]{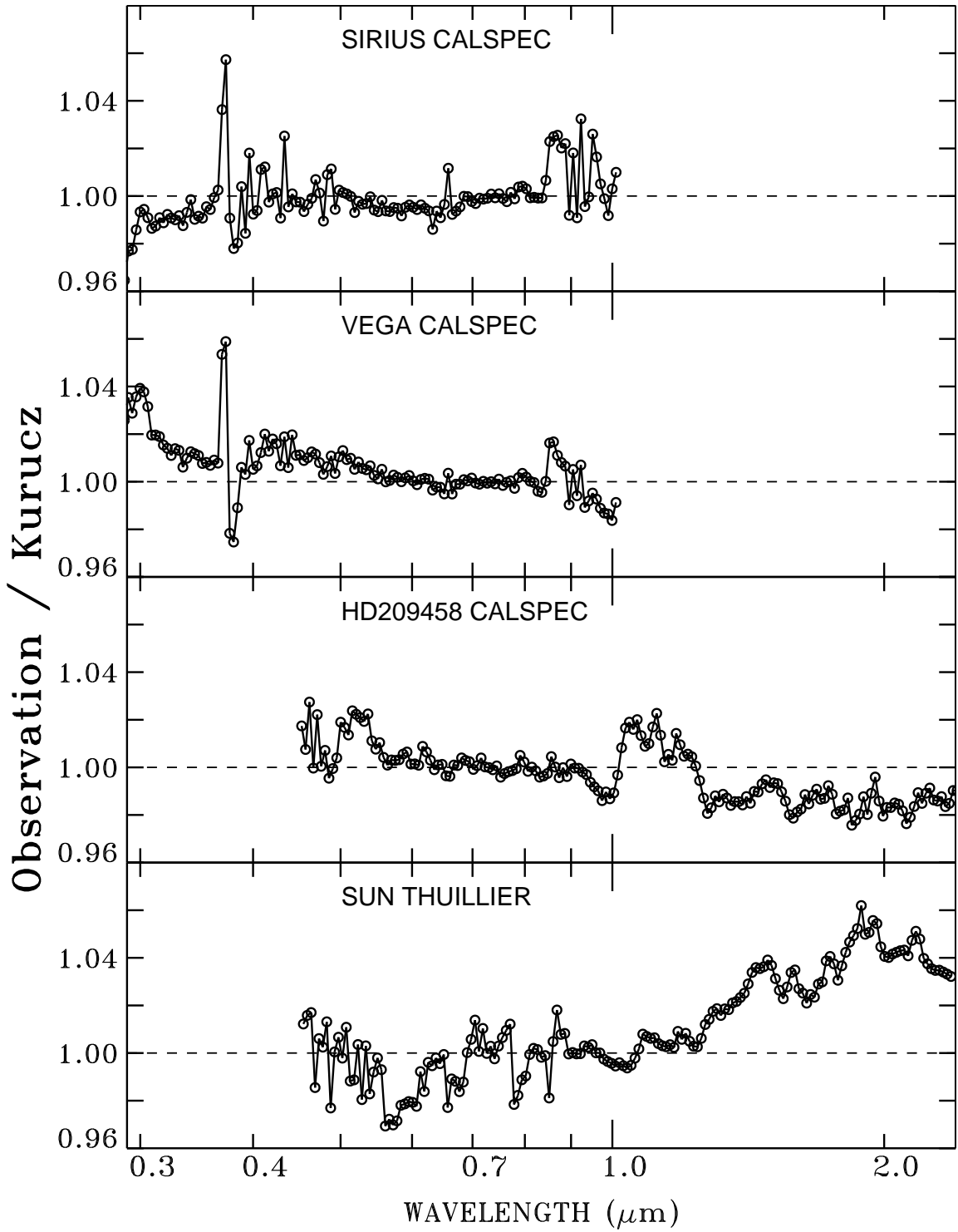}
\caption{\baselineskip=12pt
Ratio at a resolution of R=100 of measured flux to the specially tailored
Kurucz models for four stars. The models are normalized to the data at
0.7--0.8~\micron.
\label{badsun}}\end{figure}

In principle, a model atmosphere for any star can establish a primary standard
SED, just as for the standard WDs G191B2B, GD153, and GD71 in section 3.2.1
above. However, the determination of precise $T_\mathrm{eff}$ and $\log g$ from
Balmer line profiles is complicated by surface convection for G and F stars
(e.g. Ludwig et al. 2009) and by perturbing metal atoms. Modeling stellar SEDs
with significant metal line blanketing is a challenging enterprise when the goal
is an SED with 1\% relative precision over a broad wavelength region. However,
several complete SEDs for selected stars are available on the R. Kurucz
website\footnote{http://kurucz.harvard.edu}, e.g. the Sun, Vega, Sirius, and
HD209458. As shown in Figure~\ref{badsun}, the agreement of these Kurucz models
with the CALSPEC fluxes is impressive. For the A stars in the top two panels,
there is one glitch at the Balmer line convergence where the atomic physics is
not quite perfected. Vega is a rapid rotator with a pole-to-equator surface
temperature gradient (Aufdenberg et al. 2006), and the CALSPEC flux rises above
the single $T_\mathrm{eff}$=9400~K model below 0.32~\micron, because the  hotter
polar region starts to dominate the UV output. At 0.17~\micron, the true flux is
$\sim$15\% above the 9400~K single temperature model. In the IR, Vega is also a
poor standard, because a dust ring contributes significant flux starting near
the K band at 2.2\micron, where Absil et al. (2013) find a 1.3\% contribution
from the ring.

For the G stars in the bottom two panels of Figure~\ref{badsun}, the comparison
is not shown below 4500~\AA, where the heavy line blanketing limits the fidelity
of the models (cf. figure 10 of Bohlin 2007). Otherwise, the CALSPEC data for
HD209458 agrees with its model to $\sim$2\%, while the solar fluxes measured by
Th03 deviate by as much as 6\%. Thus, Figure~\ref{badsun} is evidence that
stellar absolute fluxes are known to a better accuracy than the solar absolute
flux. Because Sirius is a slow rotator with no dust ring contamination, the
Kurucz model is recommended as a primary IR SED, as suggested by Cohen et al.
(1992b) and Engelke et al. (2010). See Bohlin (2014).


\section{Spectrophotometry Archives}

\subsection{CALSPEC/HST} 

Finding charts and coordinates for many of the CALSPEC standard stars appear in 
Turnshek et al. (1990), while additional coordinates are tabulated by Bohlin et
al.(1990) and on the CALSPEC website.

The CALSPEC archive contains the spectral flux
distributions of stars with STIS or NICMOS observations, which are sometimes
supplemented by older and less reliable data from the HST Faint Object
Spectrograph (FOS) spectrograph, the
IUE satellite, or the ground-based observations of Oke (1990). Complete model
atmosphere flux distributions are also included for a few stars, while several
stars have IR extensions based on models. The {\em FITS} headers specify the source
of fluxes in the various wavelength intervals, while Bohlin et al. (2001)
illustrate the relative precision of some of the various sources of SED
measurements.

\subsection{Pickles Library}

\begin{figure} 
\centering 
\includegraphics*[height=7in]{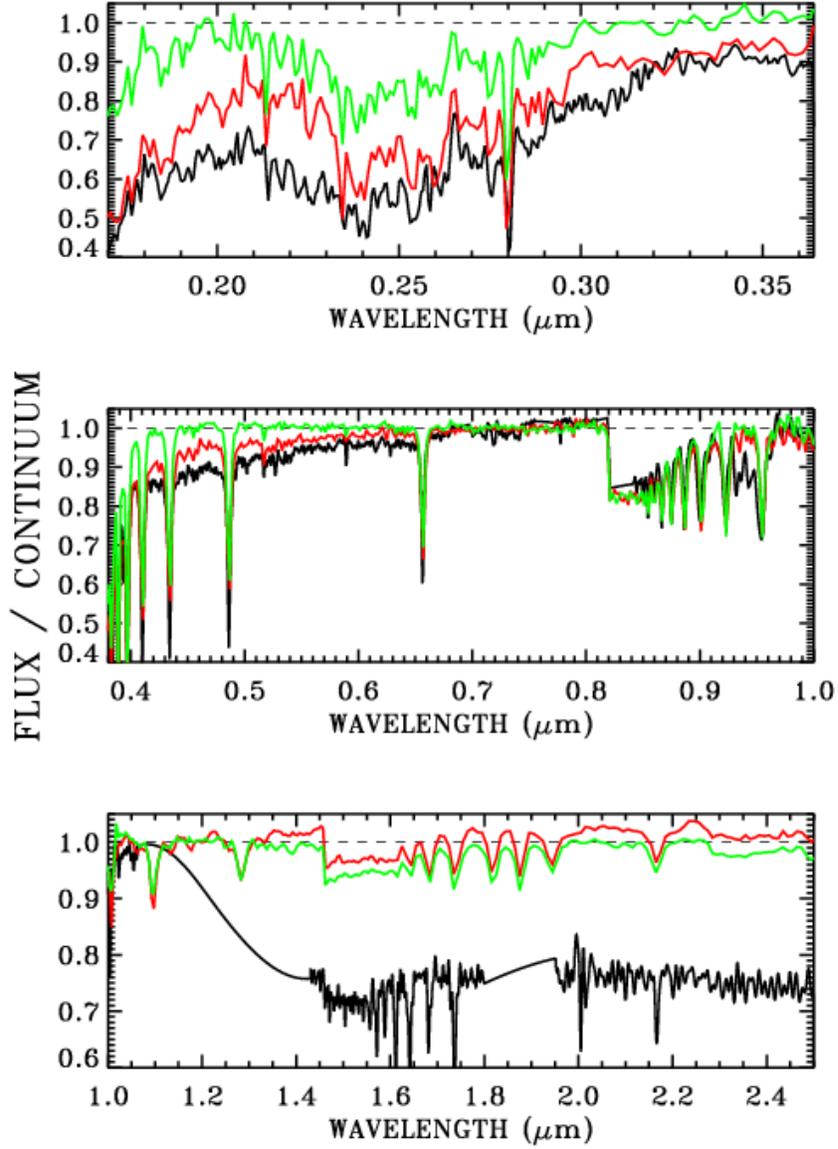}
\caption{\baselineskip=12pt
Comparison of three  stellar SEDs to the same theoretical continuum level for
$T_\mathrm{eff}$=8790~K, where the line blanketing generally increases toward
ultraviolet wavelengths. The Pickles SED for an A3V star at
$T_\mathrm{eff}$=8790~K is the black line, while the two A3V stars 1732526 (red)
and 1802271 (green) have both STIS and NICMOS data.
\label{picka}}\end{figure}

Pickles (1998) presents a 
library\footnote{http://www.ifa.hawaii.edu/users/pickles/AJP/hilib.html} of
stellar SEDs for 131 normal spectral types. The fluxes are tabulated at 5~\AA\
intervals with a resolution of $\sim$500 and cover 1150--10620~\AA\  with an
extension to 25000~\AA\ for about half of the library. The SED for each type is
constructed from a variety of sources and from observations of multiple stars of
the same spectral type. Figure~\ref{picka} compares the Pickles SED for A3V at
$T_\mathrm{eff}$=8790~K with two CALSPEC stars of the same type after
dereddening and normalizing to unity at 0.7--0.8~\micron, where the line
blanketing is minimal. These two stars are the only CALSPEC A3V spectral types
with both STIS and NICMOS measurements.

Below one micron, the three curves diverge more toward shorter wavelengths; but
differences in the effective temperature, metallicity, and reddening cause the
greatest changes in flux at far-UV wavelengths. The $T_\mathrm{eff}$=9070~K for
1802271 in Table 2 suggests that its spectral type should be somewhat hotter
than the Pickles A3 at 8790~K, which explains why the green curve is too high in
Figure~\ref{picka} in the UV. For example, a $T_\mathrm{eff}$=9070~K continuum
denominator
for 1802271 would bring the green curve down by 13\% at 0.2~\micron, while only
raising the level at 2.5~\micron\ by 3\%. For the red 1732526 curve, a
dereddening corresponding to the $E(B-V)=.061$ from Table 2 is applied; but
agreement with Pickles below one micron is within the uncertainty in the
reddening. Above one micron, the NICMOS SEDs are within a few percent of each
other, while the IR part of the Pickles SED is $\sim25\%$ too low.

\begin{figure}  
\centering  
\includegraphics*[height=7in]{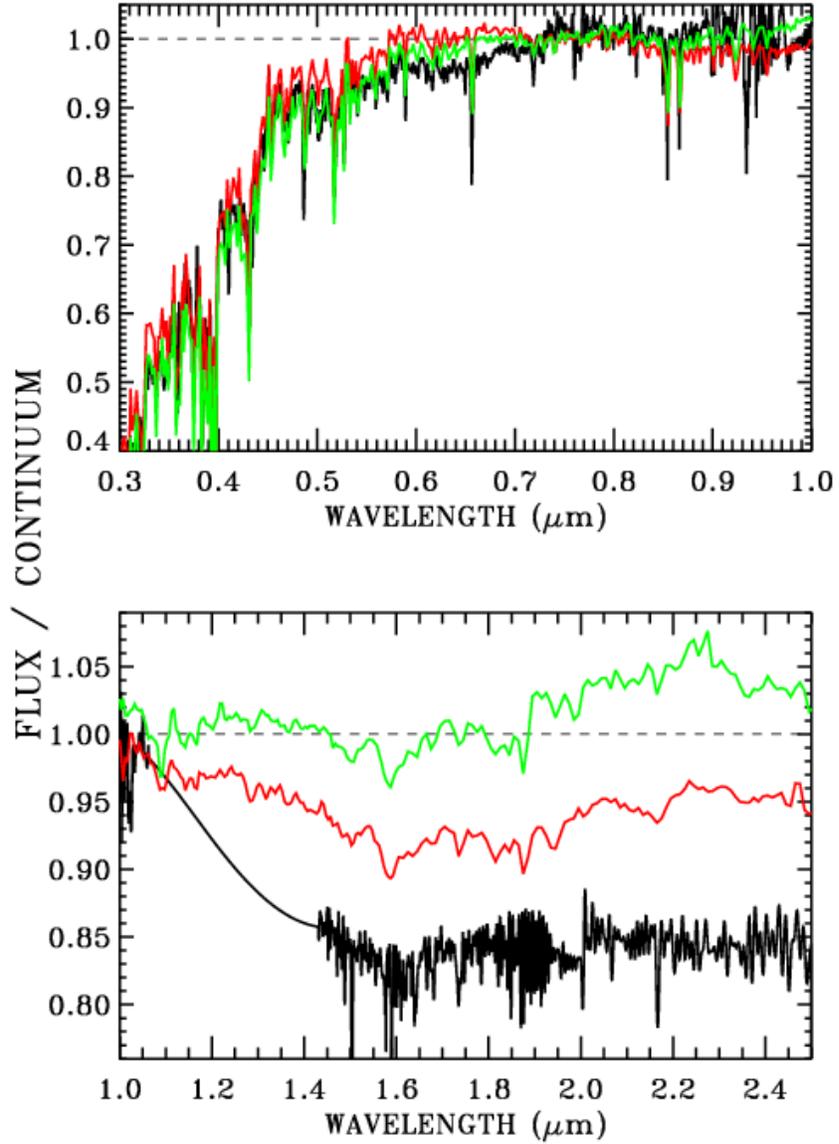}
\caption{\baselineskip=12pt 
As in Figure~\ref{picka} for G0V stars. The Pickles
SED for a G0V star at $T_\mathrm{eff}$=5800~K is the black line, while the two
CALSPEC G0V stars P041C (red) at $T_\mathrm{eff}$=6020~K and P177D (green) at
$T_\mathrm{eff}$=5880~K have STIS data below one micron and NICMOS data that
extends to 2.5\micron. \label{pickg}}\end{figure}

Figure~\ref{pickg} illustrates another comparison with Pickles at G0V and
$T_\mathrm{eff}$=5800~K for the two CALSPEC stars of the same type, viz. P041C
(6020~K) and P177D (5880~K) in Table 2. Below $\sim0.3\micron$, solar type stars
have low flux levels, which differ significantly from star-to-star. P041C is
hotter than the Pickles SED for G0V, i.e. a bit high at the shorter wavelengths
and too low in the IR. P177D (green curve) is the better reference for the
Pickles SED, which is lower than CALSPEC in the IR by $\sim15\%$.

\subsection{IUE}

The original flux calibration of the UV spectra from the IUE satellite was based
on $\eta$ UMa (Bohlin et al. 1980, Bohlin 1988). Later, Bohlin (1996) published
a correction for IUE fluxes to the same WD scale that is the basis of the HST
CALSPEC fluxes. However, the IUE project reprocessed the whole IUE archive
(Nichols \& Linsky 1996) with an optimal extraction technique (Kinney et al.
1991), which required a new flux calibration that was based entirely on the
60,000~K pure hydrogen model for the WD G191B2B. This final archive of IUE
spectra is available from the Mikulski Archive for Space Telescopes
(MAST)\footnote{http://archive.stsci.edu/iue/}.

\begin{figure}  
\centering  
\includegraphics*[trim=20 0 0 0]{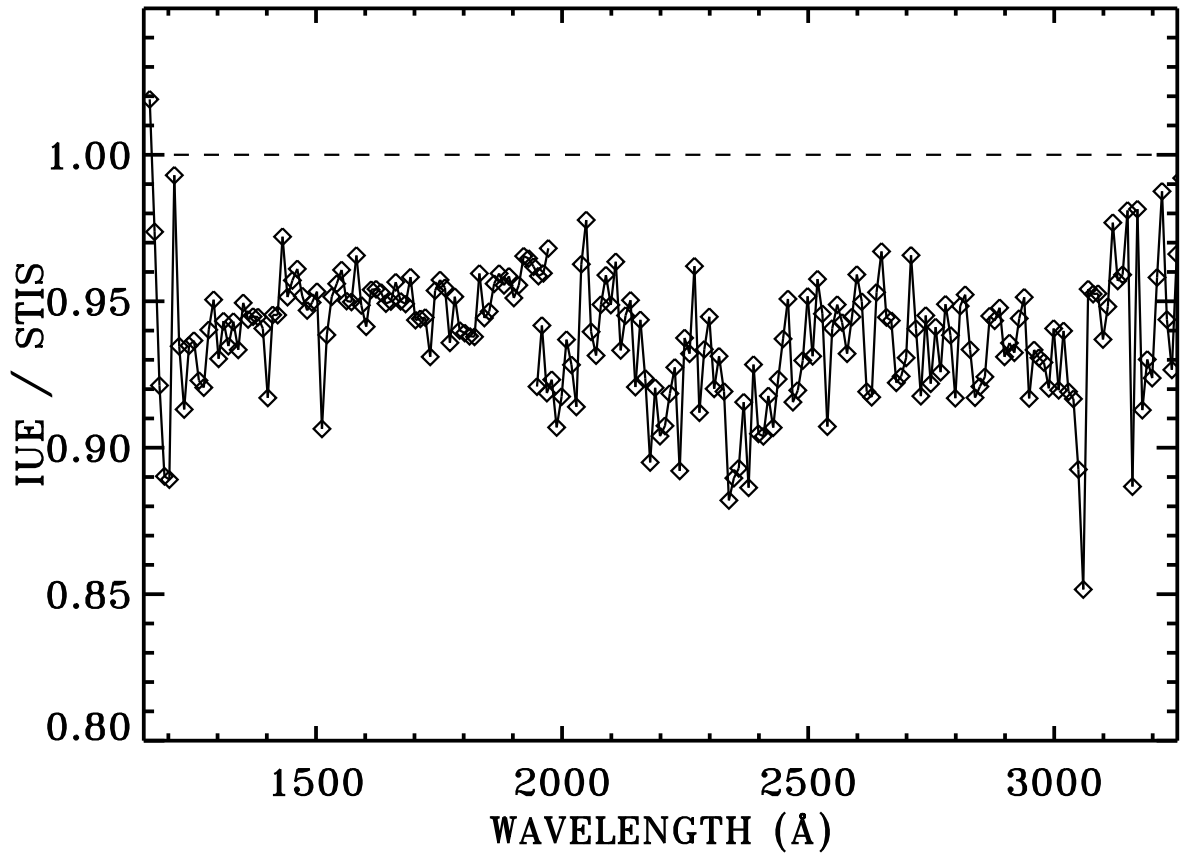}
\caption{\baselineskip=12pt
Comparison of IUE final archive fluxes and the reference standard model for
G191B2B. The ratio is in 10~\AA\ bins, and the IUE spectra are averages of 57
SWP and 48 LW individual observations of G191B2B with total exposure times of
8300 and 13500~s, respectively. \label{iue}}\end{figure}

Figure~\ref{iue} compares the IUE final archive MAST SED for G191B2B to the
current reference model from RWBK. The ratio shows little slope and is
consistent with the expectation of Nichols \& Linsky (1996). While a pure
hydrogen WD model for G191B2B provided the basis for the IUE flux calibration as
a function of wavelength, the normalization of the model was to the UV fluxes of
TD-1 (cf. Figure~\ref{figtd1} above), instead of a normalization in the visible,
e.g. the Megessier value of $F(5556)=3.46\times10^{-9}$ for Vega.
Nichols \& Linsky state that IUE
normalization is 6\% lower than the HST flux scale, which is consistent with
Figure~\ref{iue}.

\subsection{FUSE}

\begin{figure}  
\centering  
\includegraphics*[trim=20 0 0 0]{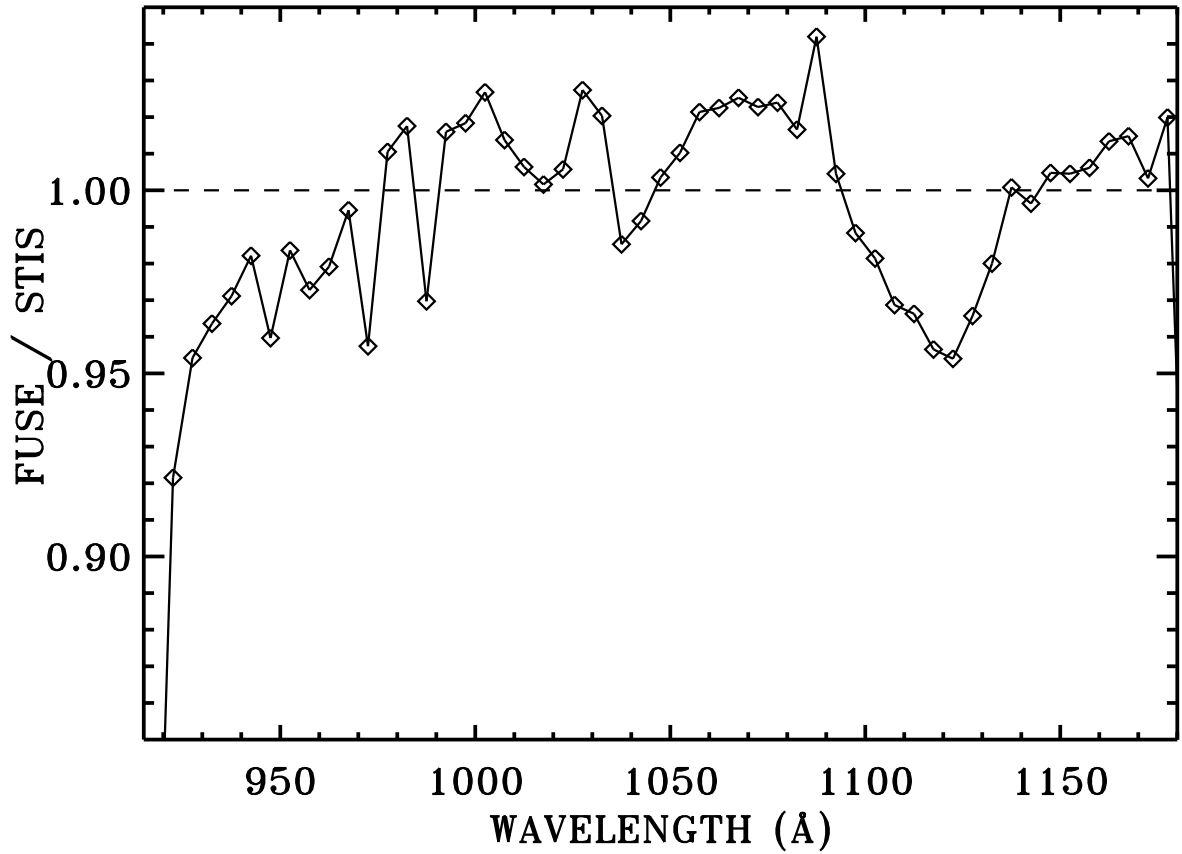}
\caption{\baselineskip=12pt
Comparison of FUSE archival fluxes with the reference standard model for
G191B2B. The ratio is in 5~\AA\ bins, and the total FUSE exposure time is
15051s. \label{fuse}}\end{figure}

The FUSE astronomical satellite recorded far-UV spectra in the 905-1187~\AA\
spectral region with a resolution of R=20,000 (Moos et al. 2000). The dynamic
range of FUSE was 10--11 magnitudes with a bright limit of V$\sim$11 for an
unreddened O star. The flux calibration is based on TLUSTY version 200 NLTE pure
hydrogen models for six WDs, including the three primary HST flux standards
(Dixon et al. 2007). The models are normalized to the V magnitude of each star.
A comparison of one FUSE observation (P104120300000nvo4histfcal.fit) for G191G2B
is compared to our Rauch model in  Figure~\ref{fuse} after correcting for the
radial velocity of 22~$km~s^{-1}$. This FUSE spectrum is one
of 61 observations of G191B2B, from the MAST
archive\footnote{http://archive.stsci.edu/fuse/}, which contains over 6000 FUSE
data sets.

\subsection{SDSS} 

\begin{figure}
\centering  
\includegraphics*[width=1.05\textwidth,trim=50 0 0 0]{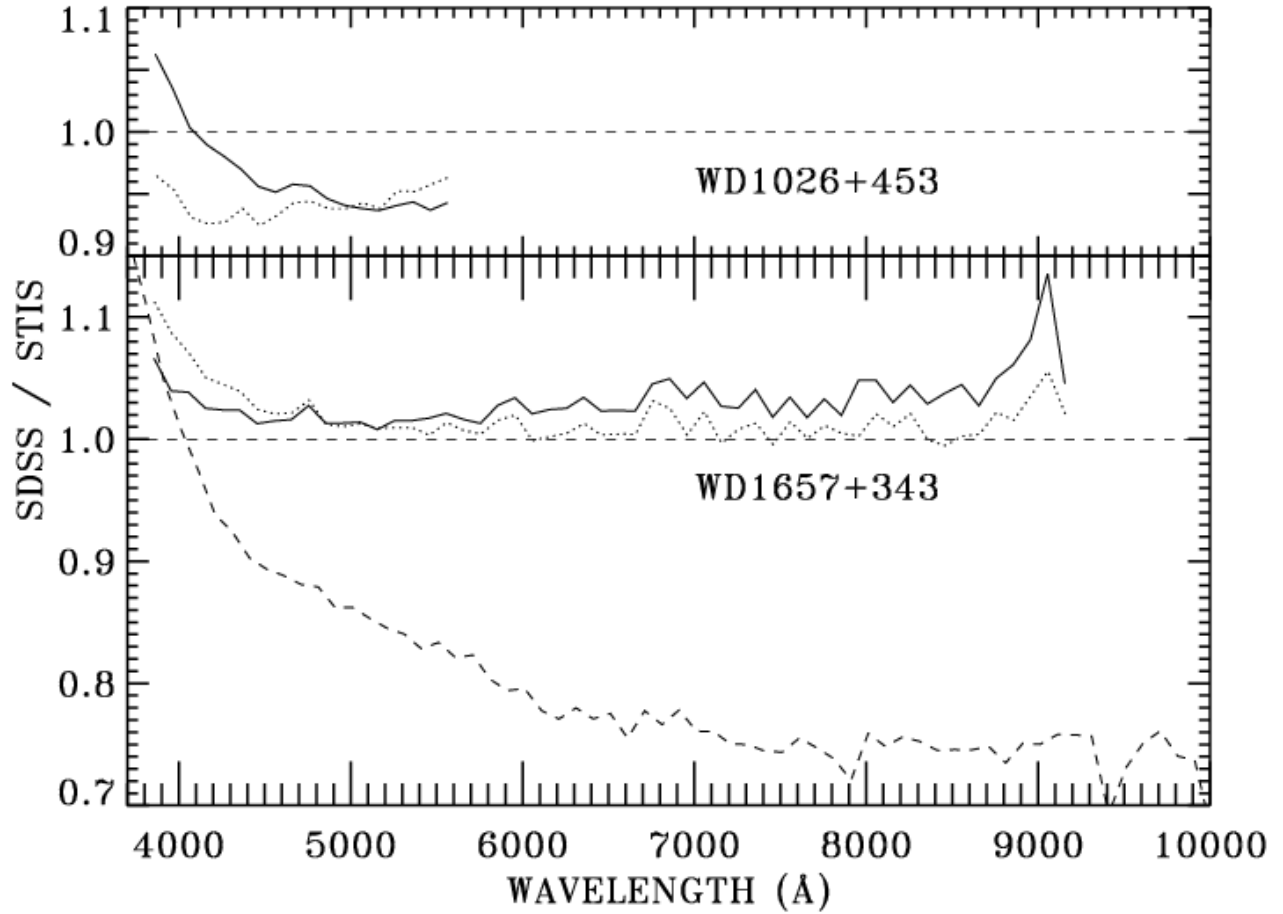}
\caption{\baselineskip=12pt
Comparison of SDSS archival fluxes with two of the faintest CALSPEC WDs in
100~\AA\ bins. Unfortunately, there are no long wavelength G750L STIS
observations for WD1026+453, so that comparison ends at 5500~\AA. The solid
and dotted lines are for early SDSS data releases, while the dashed line for
WD1657+343 is from the recent DR9 with expanded wavelength coverage.
\label{sdss}}\end{figure}

The Sloan Digital Sky Survey (SDSS) is a massive program to image a large
fraction of the northern hemisphere (Ahn et al. 2012). In conjuction with the
imaging survey, fiber spectra were obtained for about 2.6 million objects,
including $\sim$0.7 million stars in the range of g magnitude from 14 to 20
(Yanny et al. 2009).
The flux calibration was originally based on the absolute flux of
BD+17$^{\circ}$4708 (Fukugita 1996), whose flux was compared to STIS data by
Bohlin \& Gilliland (2004b). Betoule et al. (2013) discuss improvements to the
SDSS fluxes. The spectra for the two stars in common with the CALSPEC archive
are retrieved from the SDSS DR9
database\footnote{http://dr9.sdss3.org/advancedSearch} and are compared with the
CALSPEC fluxes in Figure~\ref{sdss}. The solid and dotted lines for the early
SDSS data releases are consistent with each other and with the CALSPEC fluxes
mostly within $\sim$5\%, while the newer DR9 dashed line for WD1657+343 is
discrepant by up to $\sim$25\%. A revised flux calibration for the new DR9 
instrumentation was not yet implemented.

\subsection{Infrared Space Observatory}

\begin{figure}
\centering  
\includegraphics*[width=1.05\textwidth,trim=10 0 0 0]{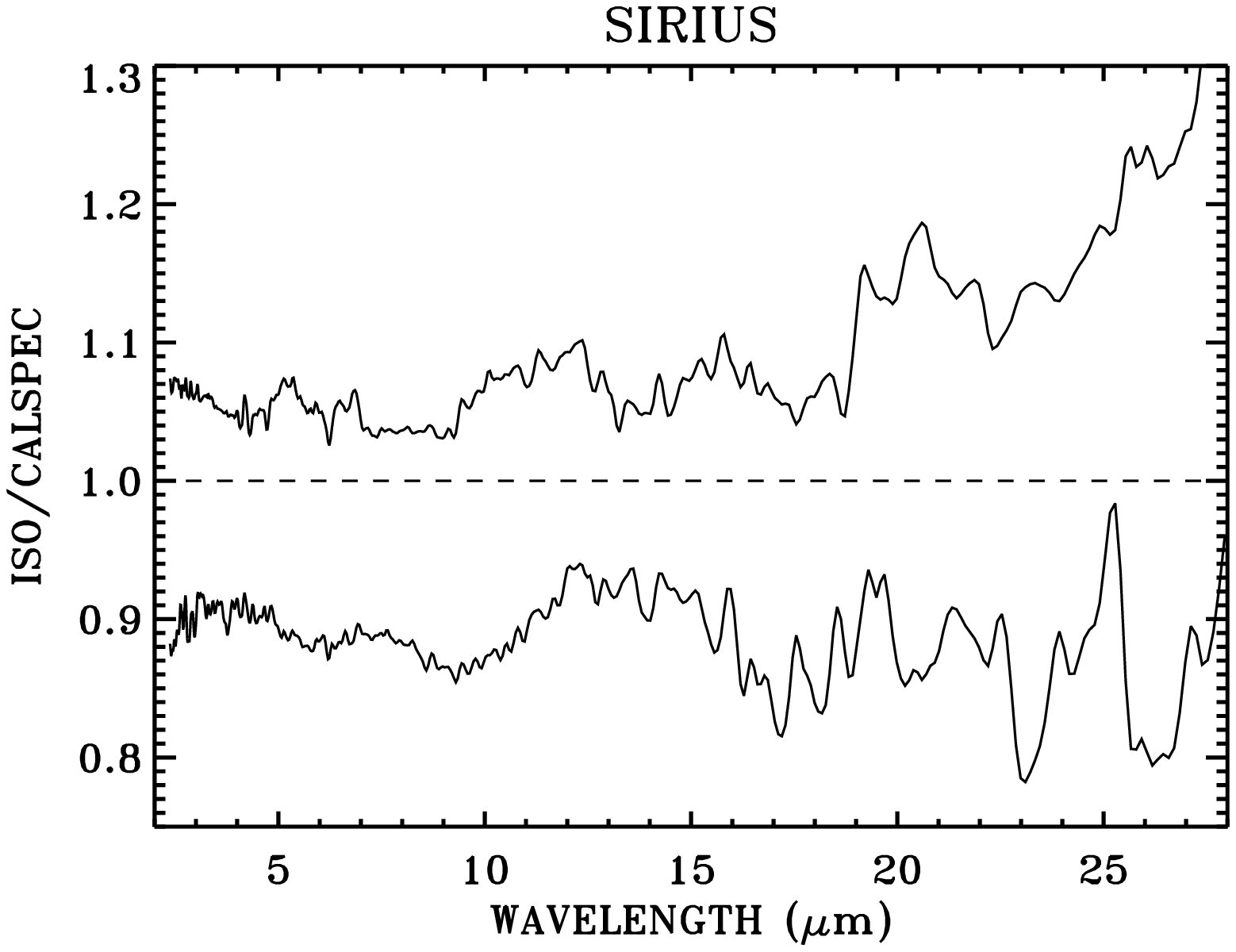}
\caption{\baselineskip=12pt
Comparison of ISO archival fluxes with the Sirius CALSPEC SED at a resolution of
R=100. The ratio above unity is for an ISO speed of 4, while the lower curve
is the ISO SED at a speed of one. \label{figiso}}\end{figure}

A comparison of the ISO short wave spectrometer
fluxes\footnote{http://isc.astro.cornell.edu/$\sim$sloan/library/2003/swsatlas.html}
for Sirius with the CALSPEC reference standard of Bohlin (2014) appears in
Figure~\ref{figiso}.  The different curves correspond to different
scan speeds with the differences reflecting different samplings of the
residual non-linearities of the detectors. If the two separate ISO
scans could be averaged, the agreement would be excellent at the few
percent level.

\begin{figure}
\centering  
\includegraphics*[width=1.05\textwidth]{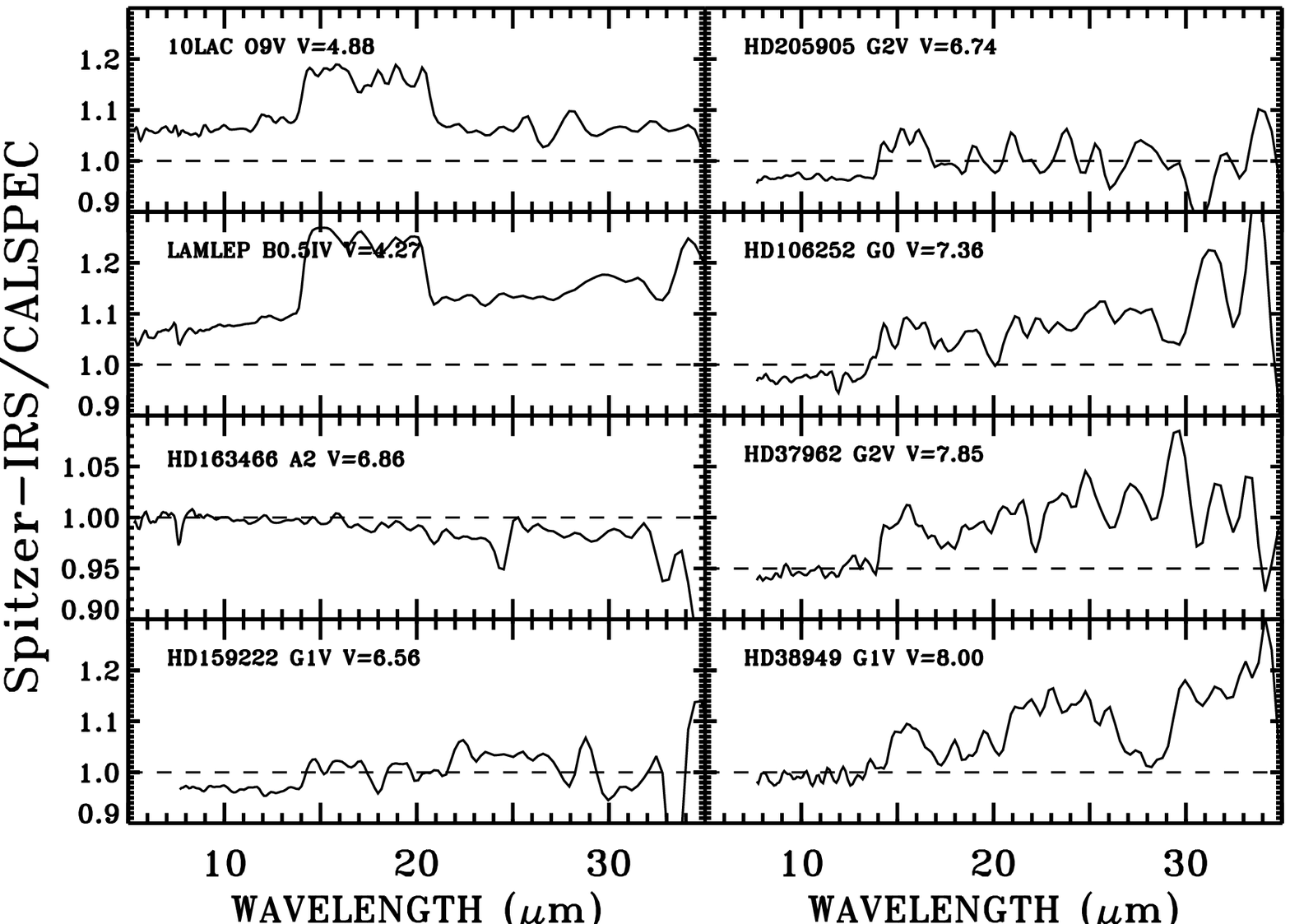}
\caption{\baselineskip=12pt
Comparison of Spitzer IRS archival fluxes with the CALSPEC SEDs at a resolution
of R=50. Notice the expanded Y-scale for HD163466, where three IRS spectra are
averaged to improve the S/N. The five G stars are arranged in decreasing
brightness down the columns; and show increasing scatter toward longer
wavelengths for fainter stars with lower signal strength.
\label{figspz}}\end{figure}

\subsection{Spitzer Space Telescope} 

Figure~\ref{figspz} compares measured spectra from the CASSIS 
reductions\footnote{http://cassis.sirtf.com/atlas/} of Spitzer Space
Telescope Infrared Spectrometer observations with the CALSPEC fluxes that are
extrapolated from the shorter wavelengths using CK04 models. One star, HD163466,
has multiple observations in the IRS archive; and the illustrated average of
three of these independent observations is on an expanded Y-scale. This average
IRS data for HD163466  agrees with CALSPEC to $<$1\% at the shorter wavelengths
and to $\sim$2\% in the 25--30~\micron\ range. Judging from the  scatter of the
other stars at the longer wavelengths, this $\sim$2\% difference for HD163466
probably reflects the statistical reproducability of the IRS data; and more IRS
scans would need to be analyzed to measure any systematic deviation of IRS
fluxes from the CALSPEC scale.

Even though the other
seven stars have only single IRS scans with correspondingly poorer S/N, the
following conclusions are suggested:

a) The O and the B stars lie considerably more above unity than the A and G
stars. So, either the hot models are less precise than the cooler part of the
CK04 grid; or there is an instrumental problem such as a shorter wavelength
light leak or contamination from a different spectral order that is significant
for the hottest stars. The elevated plateau in the 14--20~\micron\ range is
strong evidence for some sort of instrumental problem for the two hottest stars.

b) The scatter increases for the fainter stars, especially longward of
14~\micron. Thus, the best photometric comparison is below 14~\micron\ for
the six cooler stars.

c) For the A and G stars, CALSPEC and the Spitzer IRS flux shortward of
14~\micron\ agree to better than 3\%, which suggest that neither estimate has
systematic errors of more than $\sim$3\% in absolute flux.





\section{Future Directions}

On the theoretical front, disparities in the NLTE modeling is the largest 
limitation to the precision of the three primary WD flux standards. Ideally, the
pre-eminent developers of the modeling codes might confer and determine whether
the discrepancies in SEDs calculated for the same  $T_\mathrm{eff}$ and $\log g$
are caused by different input physics or numerical imprecisions. Secondarily,
the uncertainties on the derivation of $T_\mathrm{eff}$ and $\log g$ are a floor
to the precision of the models (dotted line in Figure~\ref{moderr}). Perhaps,
better data or improved analysis techniques could reduce this contribution of
the systematic error budget of the fundamental WD models.

Experimentally, the NIST laboratory reference standards that have sub-percent
precision should be transferred to the stars. Rocket or other space platforms
provide the most straightforward path for implementation. However, the extensive
programs to measure the temporal and wavelength variations of the atmospheric
transmission in the visible and near-IR may well produce stellar standards with
precise flux distributions at the top of the atmosphere. To verify the MSX
results in the mid-IR and extend the MSX concept to other wavelengths, a new
mission is needed with on-board, in-situ absolute flux sources like the MSX
ejected reference spheres. The success of these experimental programs would free
the flux standards from the uncertainties of the theoretical model calculations.

In summary, the current set of CALSPEC stars are the best available stellar flux
standards for the FUV to mid-IR, where the SEDs as a function of wavelength are
based on NLTE models atmospheres for the primary standard WD stars of Table 1.
Table 2 includes many of the secondary CALSPEC standards.

\acknowledgments

Thanks to Derck Massa for a patient explanation of the nuances of covariance
matrices. S. Deustua, M. Kaiser, R. Kurucz, and the referee provided
constructive comments on preliminary drafts. Support for this work was provided
by NASA through the Space Telescope Science Institute, which is operated by
AURA, Inc., under NASA contract NAS5-26555. This research made use of the SIMBAD
database, operated at CDS, Strasbourg, France.

\end{document}